\documentclass[11pt]{article}
\usepackage{fullpage, verbatim}
\usepackage{amsmath, amsthm, amssymb}

\author{ Mihai P\v{a}tra\c{s}cu \\ {\tt mip@mit.edu}
  \and   Mikkel Thorup \\ {\tt mthorup@research.att.com} }

\title{ Time-Space Trade-Offs for Predecessor Search%
  \footnote{An extended abstract of this paper appears in the
    Proceedings of the 38th ACM Symposium on Theory of Computing
    (STOC'06).}}

\newtheorem{lemma}{Lemma}
\newtheorem{theorem}[lemma]{Theorem}
\newtheorem{proposition}[lemma]{Proposition}

\let\latexcite=\cite
\def\cite{\nolinebreak\latexcite}
\let\latexref=\ref
\def\ref{\nolinebreak\latexref}

\newcommand{\req}[1]{(\ref{eq:#1})}    

\newcommand{\eps}{\varepsilon}
\renewcommand{\emptyset}{\varnothing}

\newcommand{\E}{\mathbf{E}}            

\newcommand{\func}[1] {\textnormal{\tt\scshape#1}}

\newcommand{\dsum}[2]{\bigoplus^{#1} {#2}}
\newcommand{\fold}[2]{{#1}^{({#2})}}

\newcommand{\Distr}{\mathcal{D}}       
\newcommand{\Fam}{\mathcal{F}}         
\newcommand{\cells}{\Gamma}            
\newcommand{\kl}{\ell}                 
\newcommand{\ws}{w}                    

\newcommand{\randd}{\mathbf{d}}        
\newcommand{\randr}{\mathbf{r}}        
\newcommand{\randQ}{\mathbf{Q}}        
\newcommand{\randp}{\mathbf{p}}        
\newcommand{\randi}{\mathbf{i}}        
\newcommand{\randq}{\mathbf{q}}        

\newcommand{\fixr}{\mathfrak{r}}
\newcommand{\fixQ}{\mathfrak{Q}}

\def\concat{\cdot}
\def\fmax{\textit{max}}
\def\fSuff{\textit{Suff}}
\def\fNextchar{\textit{Next\_\,char}}
\def\fAgree{\textit{Agree}}
\def\fcompref{\textit{comm\_\,pref}}
\def\fPred{\textit{Pred}}
\def\fpred{\textit{pred}}
\def\fspred{\textit{pred}^{\,\neq}}
\def\fprefix{\textit{prefix}}
\def\fsuffix{\textit{suffix}}

\begin{document}

\maketitle

\setcounter{page}{0}
\thispagestyle{empty}

\begin{abstract}
We develop a new technique for proving cell-probe lower bounds for
static data structures. Previous lower bounds used a reduction to
communication games, which was known not to be tight by counting
arguments. We give the first lower bound for an explicit problem which
breaks this communication complexity barrier. In addition, our bounds
give the first separation between polynomial and near linear
space. Such a separation is inherently impossible by communication
complexity.

Using our lower bound technique and new upper bound constructions, we
obtain tight bounds for searching predecessors among a static set of
integers. Given a set $Y$ of $n$ integers of $\kl$ bits each, the goal
is to efficiently find $\func{predecessor}(x) = \max~\{ y \in Y \mid y
\le x \}$. For this purpose, we represent $Y$ on a RAM with word
length $\ws$ using $S$ words of space. Defining $a = \lg \frac{S}{n} +
\lg \ws$, we show that the optimal search time is, up to constant
factors:
\begin{equation*}
\min \left\{ \begin{array}{l}
  \log_{\ws} n \\[1.5ex]
  \lg \frac{\kl - \lg n}{a} \\[1.5ex]
  \frac{\lg \frac{\kl}{a}}{\lg \left( \frac{a}{\lg n} 
    \,\cdot\, \lg \frac{\kl}{a} \right)} \\[2.5ex]
  \frac{\lg \frac{\kl}{a}}{\lg \left( \lg \frac{\kl}{a} 
    \textrm{\large~/} \lg\frac{\lg n}{a} \right)}
\end{array} \right.
\end{equation*}

In external memory ($\ws > \kl$), it follows that the optimal strategy
is to use either standard B-trees, or a RAM algorithm ignoring the
larger block size. In the important case of $\ws = \kl = \gamma \lg
n$, for $\gamma > 1$ (i.e.~polynomial universes), and near linear
space (such as $S = n \cdot \lg^{O(1)} n$), the optimal search time is
$\Theta(\lg \kl)$. Thus, our lower bound implies the surprising
conclusion that van Emde Boas' classic data structure from [FOCS'75]
is optimal in this case. Note that for space $n^{1+\eps}$, a running
time of $O(\lg \kl / \lg\lg \kl)$ was given by Beame and Fich
[STOC'99].
\end{abstract}

\clearpage

\section{Introduction}

In this paper we provide tight trade-offs between query time and space
of representation for static predecessor search. This is one of the
most basic data structures, and the trade-off gives the first
separation between linear and polynomial space for {\em any\/} data
structure problem.

\subsection{The Complexity-Theoretic View}

Yao's cell-probe model \cite{yao81tables} is typically the model of
choice for proving lower bounds on data structures. The model assumes
the memory is organized in $\ws$-bit cells (alternatively called
words). In the case of static data structures, one first constructs a
representation of the input in a table with a bounded number of cells
$S$ (the space complexity). Then, a query can be answered by probing
certain cells. The time complexity $T$ is defined to be the number of
cell probes. The model allows free nonuniform computation for both
constructing the input representation, and for the query
algorithm. Thus, the model is stronger than the word RAM or its
variants, which are used for upper bounds, implementable in a
programming language like C. In keeping with the standard assumptions
on the upper bound side, we only consider $\ws = \Omega(\lg n)$.

Typically, lower bounds in this model are proved by considering a
two-party communication game. Assume Bob holds the data structure's
input, while Alice holds the query. By simulating the cell-probe
solution, one can obtain a protocol with $T$ rounds, in which Alice
sends $\lg S$ bits and Bob replies with $\ws$ bits per round. Thus, a
lower bound on the number of rounds translates into a cell-probe lower
bound.

Intuitively, we do not expect this relation between cell-probe and
communication complexity to be tight. In the communication model, Bob
can remember past communication, and answer new queries based on
this. Needless to say, if Bob is just a table of cells, he cannot
remember anything, and his responses must be a function of Alice's
last message (i.e.~the address of the cell probe). By counting
arguments, it can be shown \cite{miltersen93bitprobe} that the
cell-probe complexity can be much higher than the communication
complexity, for natural ranges of parameters. However, a separation
for an explicit problem has only been obtained in a very restricted
setting. G\'{a}l and Miltersen \cite{gal03succinct} showed such a
separation when the space complexity is very close to minimum: given
an input of $n$ cells, the space used by the data structure is $n +
o(n)$.

Besides the reduction to communication complexity, and the approach of
\cite{gal03succinct} for very small space, there are no known
techniques applicable to static cell-probe complexity with cells of
$\Omega(\lg n)$ bits. In particular, we note that the large body of
work initiated by Fredman and Saks \cite{fredman89cellprobe} only
applies to \emph{dynamic} problems, such as maintaining partial sums
or connectivity. In the case of static complexity, there are a few
other approaches developed specifically for the bit-probe model ($\ws =
1$); see \cite{miltersen99survey}.

In conclusion, known lower bound techniques for cell-probe complexity
cannot surpass the communication barrier.  However, one could still
hope that communication bounds are interesting enough for natural data
structure problems. Unfortunately, this is often not the case. Notice
that polynomial differences in $S$ only translate into constant
factors in Alice's message size. In the communication game model, this
can only change constant factors in the number of rounds, since Alice
can break a longer message into a few separate messages.
Unfortunately, this means that communication complexity cannot be used
to separate, say, polynomial and linear space. For many natural
data-structure problems, the most interesting behavior occurs close to
linear space, so it is not surprising that our understanding of static
data-structure problems is rather limited.

In this work, we develop a new lower-bound technique, the
\emph{cell-probe elimination lemma}, targeted specifically at the
cell-probe model. Using this lemma, we obtain a separation between
space $n^{1+o(1)}$ and space $n^{1+\eps}$ for any $\eps > 0$. This
also represents a separation between communication complexity and
cell-probe complexity with space $n^{1+o(1)}$. Our lower bounds hold
for predecessor search, one of the most natural and well-studied
problems.

Our lower bound result has a strong direct sum flavor, which is
interesting in its own right. Essentially, we show that for problems
with a certain structure, a data structure solving $k$ independent
subproblems with space $k\cdot \sigma$ cannot do better than $k$ data
structures solving each problem with space $\sigma$.

\subsection{The Data-Structural View}

Using our lower bound technique and new upper bound constructions, we
obtain tight bounds for predecessor search. The problem is to
represent an ordered set $Y$, such that for any query $x$ we can find
efficiently $\func{predecessor}(x) = \max~\{ y \in Y \mid y \le x \}$.
This is one of the most fundamental and well-studied problems in data
structures. For a comprehensive list of references, we refer to
\cite{beame02pred}; here, we only describe briefly the best known
bounds.

\subsubsection{The Upper-Bound Story}

We focus on the static case, where $Y$ is given in advance for
preprocessing. For example, we can sort $Y$, and later find the
predecessor of $x$ by binary search using $O(\lg n)$ comparisons,
where $n=|Y|$.

On computers, we are particularly interested in integer keys. Thereby
we also handle, say, floating point numbers whose ordering is
preserved if they are cast as integers. We can then use all the
instructions on integers available in a standard programming language
such as C, and we are no longer limited by the $\Omega(\lg n)$
comparison based lower bound for searching. A strong motivation for
considering integer keys is that integer predecessor search is
asymptotically equivalent to the IP look-up problem for forwarding
packets on the Internet \cite{feldmann00ip}. This problem is extremely
relevant from a practical perspective. The fastest deployed software
solutions use non-comparison-based RAM tricks \cite{degermark97route}.

More formally, we will represent $Y$ on a unit-cost word RAM with a
given word length $\ws$. We assume each integers in $Y$ has $\kl$
bits, and that $\lg n \le \kl \le \ws$. On the RAM, the most natural
assumption is $\kl = \ws$. The case $\ws > \kl$ models the external
memory model with $B = \lfloor \frac{\ws}{\kl} \lfloor$ keys per
page. In this case, the well-known (comparison-based) B-trees achieve
a search time of $O(\log_B n)$. For the rest of the discussion, assume
$\ws = \kl$.

Using the classic data structure of van Emde Boas \cite{vEB77pred}
from 1975, we can represent our integers so that predecessors can be
searched in $O(\lg \kl)$ time. The space is linear if we use hashing
\cite{willard83pred}.

In the 1990, Fredman and Willard \cite{fredman93fusion} introduced
fusion trees, which requires linear space and can answer queries in
$O(\log_{\kl} n)$ time. Combining with van Emde Boas' data
structure, they got a search time of $O(\min\, \{ \frac{\lg n}{\lg
\kl},\, \lg \kl \})$, which is always $O(\sqrt{\lg n})$.

In 1999, Beame and Fich \cite{beame02pred} found an improvement to van
Emde Boas' data structure bringing the search time down to
$O(\frac{\lg \kl} {\lg\lg \kl})$. Combined with fusion trees, this
gave them a bound of $O(\min\, \{ \frac{\lg n}{\lg \kl},\, \frac{\lg
\kl} {\lg\lg \kl} \})$, which is always $O(\sqrt{\frac{\lg n}{\lg\lg
n}})$. However, the new data structure of Beame and Fich uses
quadratic space, and they asked if the space could be improved to
linear or near-linear.

As a partially affirmative answer to this question, we show that their
$O( \frac{\lg \kl}{\lg\lg \kl} )$ search time can be obtained with
space $n^{1 + 1 / \exp(\lg^{1-\eps} \kl)}$ for any $\eps > 0$.
However, we also show, as our main result, that with closer to linear
space, such as $n\lg^{O(1)} n$, one cannot in general improve the old
van Emde Boas bound of $O(\lg \kl)$.

\subsubsection{The Lower-Bound Story}

Ajtai \cite{ajtai88pred} was the first to prove a superconstant lower
bound for our problem. His results, with a correction by Miltersen
\cite{miltersen94pred}, can be interpreted as saying that there exists
$n$ as a function of $\kl$ such that the time complexity for
polynomial space is $\Omega(\sqrt{\lg \kl})$, and likewise there
exists $\kl$ a function of $n$ making the time complexity
$\Omega(\sqrt[3]{\lg n})$.

Miltersen \cite{miltersen94pred} revisited Ajtai's work, showing that
the lower bound holds in the communication game model, and for a
simpler colored predecessor problem. In this problem, the elements of
$Y$ have an associated color (say, red or blue), and the query asks
only for the color of the predecessor in $Y$. This distinction is
important, as one can reduce other problems to this simpler problem,
such as existential range queries in two dimensions
\cite{miltersen99asymmetric} or prefix problems in a certain class of
monoids \cite{miltersen94pred}. Like previous lower bound proofs, ours
also holds for the colored problem, making the lower bounds applicable
to these problems.

Miltersen, Nisan, Safra and Wigderson \cite{miltersen99asymmetric}
once again revisited Ajtai's proof, extending it to randomized
algorithms. More importantly, they captured the essence of the proof
in an independent \emph{round elimination lemma}, which forms a
general tool for proving communication lower bounds. Our cell-probe
elimination lemma is inspired, at a high level, by this result.

Beame and Fich \cite{beame02pred} improved the lower bounds to
$\Omega(\frac{\lg \kl}{\lg\lg \kl})$ and $\Omega(\sqrt{\frac{\lg n}
{\lg\lg n}})$ respectively. Sen and Venkatesh \cite{sen-roundelim}
later gave an improved round elimination lemma, which can reprove the
lower bounds of Beame and Fich, but also for randomized algorithms.
Analyzing the time-space trade-offs obtained by these proofs, one
obtains $\Omega( \frac{\lg n}{\lg \ws}, \frac{\lg \kl}{\lg\lg S})$,
where $S$ is the space bound, and possibly $\ws > \kl$.

\subsection{The Optimal Trade-Offs}

Define $\lg x = \lceil \log_2 (x+2) \rceil$, so that $\lg x \geq 1$
even if $x \in [0,1]$. Assuming space $S$, and defining $a = \lg
\frac{S}{n} + \lg \ws$, we show that the optimal search time is, up to
constant factors:
\begin{equation}  \label{eq:full-bound}
\min \left\{ \begin{array}{l}
  \log_{\ws} n \\[1.5ex]
  \lg \frac{\kl - \lg n}{a} \\[1.5ex]
  \frac{\lg \frac{\kl}{a}}{\lg \left( \frac{a}{\lg n} 
    \,\cdot\, \lg \frac{\kl}{a} \right)} \\[2.5ex]
  \frac{\lg \frac{\kl}{a}}{\lg \left( \lg \frac{\kl}{a} 
    \textrm{\large~/} \lg\frac{\lg n}{a} \right)}
\end{array} \right.
\end{equation}

The upper bounds are achieved by a deterministic query algorithm on a
RAM. The data structure can be constructed in expected time $O(S)$ by
a randomized algorithm, starting from a sorted list of integers. The
lower bounds hold for deterministic query algorithms answering the
colored predecessor problem in the cell-probe model. When $S \ge
n^{1+\eps}$ for some constant $\eps > 0$, the lower bounds also hold
in the stronger communication game model, even allowing randomization
with two-sided error.

\subsubsection{External Memory and Branch One}

To understand the first branch of the trade-off, first consider the
typical case on a RAM, when a word fits exactly one integer, i.e.~$\ws
= \kl$. In this case, the bound is $\log_\kl n$, which describes the
performance of fusion trees \cite{fredman93fusion}.

To understand the case $\ws > \kl$, consider the external memory model
with $B$ words per page. This model has as a nonuniform counterpart
the cell-probe model with cells of size $\ws = B\kl$. Observe that
only the first branch of our trade-off depends on $\ws$. This branch
is $\log_\ws n = \frac{\lg n}{\lg B + \lg \kl} = \Theta(\min \{
\log_\kl n, \log_B n \} )$. The first term describes the performance
of fusion trees on a RAM with $\kl$-bit words, as noted above. The
second term matches the performance of the B-tree, the fundamental
data structure in external memory.

Thus, we show that it is always optimal to either use a standard
B-tree, or the best RAM algorithm which completely ignores the
benefits of external memory. The RAM algorithm uses $\kl$-bit words,
and ignores the grouping of words into pages; this algorithm is the
best of fusion trees and the algorithms from branches 2--4 of the
trade-off. Thus, the standard comparison-based B-tree is the optimal
use of external memory, even in a strong model of computation.

\subsubsection{Polynomial Universes: Branch Two}

For the rest of the discussion, assume the first branch (B-trees and
fusion trees) does not give the minimum. Some of the most interesting
consequences of our results can be seen in the very important special
case when integers come from a polynomial universe, i.e.~$\kl = O(\lg
n)$. In this case, the optimal complexity is $\Theta(\lg \frac{\kl -
\lg n}{a})$, as given by the second branch of the trade-off. 

On the upper bound side, this is achieved by a simple elaboration of
van Emde Boas' data structure. This data structure gives a way to
reduce the key length from $\kl$ to $\frac{\kl}{2}$ in constant time,
which immediately implies an upper bound of $O(\lg \kl)$. To improve
that, first note that when $\kl \le a$, we can stop the recursion and
use complete tabulation to find the result. This means only $O(\lg
\frac{\kl}{a})$ steps are needed. Another trivial idea, useful for
near-linear universes, is to start with a table lookup based on the
first $\lg n$ bits of the key, which requires linear space. Then,
continue to apply van Emde Boas for keys of $w - \lg n$ bits inside
each subproblem, giving a complexity of $O(\lg \frac{w - \lg n}{a})$.

Quite surprisingly, our lower bound shows that van Emde Boas' classic
data structure, with these trivial tweaks, is optimal. In particular,
when the space is not too far from linear (at most $n \cdot
2^{\lg^{1-\eps} n}$) and $\kl \ge (1+\eps) \lg n$, the standard van
Emde Boas bound of $\Theta(\lg \kl)$ is optimal. It was often
conjectured that this bound could be improved.

Note that with space $n^{1+\eps}$, the optimal complexity for
polynomial universes is constant. However, with space $n^{1+o(1)}$,
the bound is $\omega(1)$, showing the claimed complexity-theoretic
separations.

\subsubsection{The Last Two Branches}

The last two branches are relevant for superpolynomial universes,
i.e.~$\kl = \omega(\lg n)$. Comparing the two branches, we see the
third one is better than the last one (up to constants) when $a =
\Omega(\lg n)$. On the other hand, the last branch can be
asymptotically better when $a = o(\lg n)$.  This bound has the
advantage that in the logarithm in the denominator, the factor
$\frac{a}{\lg n}$, which is subconstant for $a=o(\lg n)$, is replaced
by $1 / \lg \frac{a}{\lg n}$. 

The third branch is obtained by a careful application of the
techniques of Beame and Fich \cite{beame02pred}, which can improve
over van Emde Boas, but need large space. The last branch is also
based on these techniques, combined with novel approaches tailored for
small space.

\subsubsection{Dynamic Updates}

Lower bounds for near-linear space easily translate into interesting
lower bounds for dynamic problems. If inserting an element takes time
$t_u$, we can obtain a static data structure using space $O(n \cdot
t_u)$ by simply simulating $n$ inserts and storing the modified cells
in a hash table. This transformation works even if updates are
randomized, but, as before, we require that queries be
deterministic. This model of randomized updates and deterministic
queries is standard for hashing-based data structures.  By the
discussion above, as long as updates are reasonably fast, one cannot
in general improve on the $O(\lg \kl)$ query time.  It should be noted
that van Emde Boas data structure can handle updates in the same time
as queries, so this classic data structure is also optimal in the
typical dynamic case, when one is concerned with the slowest
operation.

\subsection{Contributions}

We now discuss our contributions in establishing the tight results of
\req{full-bound}. Our main result is proving the tight lower bounds
for $a = o(\lg n)$ (in particular, branches two and four of the
trade-off). As mentioned already, previous techniques were helpless,
since none could even differentiate $a = 2$ from $a = \lg n$.

Interestingly, we also show improved lower bounds for the case $a =
\Omega(\lg n)$, in the classic communication framework. These
improvements are relevant to the third branch of the
trade-off. Assuming for simplicity that $a \le w^{1 - \eps}$, our
bound is $\min\, \{ \frac{\lg n}{\lg w}, \frac{\lg w}{\lg\lg w + \lg
(a/\lg n)} \}$, whereas the best previous lower bound was $\min\, \{
\frac{\lg n}{\lg w}, \frac{\lg w}{\lg a} \}$. Our improved bound is
based on a simple, yet interesting twist: instead of using the round
elimination lemma alone, we show how to combine it with the
\emph{message compression lemma} of Chakrabarti and Regev
\cite{chakrabarti04ann}. Message compression is a refinement of round
elimination, introduced by \cite{chakrabarti04ann} to prove a lower
bound for the approximate nearest neighbor problem. Sen and Venkatesh
\cite{sen-roundelim} asked whether message compression is really
needed, or one could just use standard round elimination. Our result
sheds an interesting light on this issue, as it shows message
compression is even useful for classic predecessor lower bounds.

On the upper bound side, we only need to show the last two branches of
the trade-off. As mentioned already, we use techniques of Beame and
Fich \cite{beame02pred}.  The third bound was anticipated%
\footnote{As a remark in \cite[Section 7.5]{thorup03stab}, it is
  stated that ``it \emph{appears} that we can get the following
  results\dots'', followed by bounds equivalent to the third branch
  of \req{full-bound}.}
by the second author in the concluding remarks of \cite{thorup03stab}.
The last branch of \req{full-bound}, tailored specifically for small
space, is based on novel ideas.

\subsection{Direct-Sum Interpretations}

A very strong consequence of our proofs is the idea that sharing
between subproblems does not help for predecessor search. Formally,
the best cell-probe complexity achievable by a data structure
representing $k$ independent subproblems (with the same parameters) in
space $k\cdot \sigma$ is asymptotically equal to the best complexity
achievable by a data structure for one subproblem, which uses space
$\sigma$. The simplicity and strength of this statement make it
interesting from both the data-structural and complexity-theoretic
perspectives.

At a high level, it is precisely this sort of direct-sum property that
enables us to beat communication complexity. Say we have $k$
independent subproblems, and total space $S$. While in the
communication game Alice sends $\lg S$ bits per round, our results
intuitively state that $\lg \frac{S}{k}$ bits are sufficient. Then, by
carefully controlling the increase in $k$ and the decrease in key
length (the query size), we can prevent Alice from communicating her
entire input over a superconstant number of rounds.

A nice illustration of the strength of our result are the tight bounds
for near linear universes, i.e.~$\kl = \lg n + \delta$, with $\delta =
o(\lg n)$. On the upper bound side, the algorithm can just start by a
table lookup based on the first $\lg n$ bits of the key, which
requires linear space. Then, it continues to apply van Emde Boas for
$\delta$-bit keys inside each subproblem, which gives a complexity of
$O(\lg \frac{\delta}{a})$. Obtaining a lower bound is just as easy,
given our techniques. We first consider $n / 2^{\delta}$ independent
subproblems, where each has $2^{\delta}$ integers of $2\delta$ bits
each. Then, we prefix the integers in each subproblem by the number of
the subproblem (taking $\lg n - \delta$ bits), and prefix the query
with a random subproblem number. Because the universe of each
subproblem ($2^{2\delta}$) is quadratically bigger than the number of
keys, we can apply the usual proof showing the optimality of van Emde
Boas' bound for polynomial universes. Thus, the complexity is
$\Omega(\lg \frac{\delta}{a})$.

\section{Lower Bounds for Small Space}  \label{sec:smallspace}

\subsection{The Cell-Probe Elimination Lemma}   \label{sec:cpe-state}

An abstract decision data structure problem is defined by a function
$f : D \times Q \to \{0,1\}$. An input from $D$ is given at
preprocessing time, and the data structure must store a representation
of it in some bounded space. An input from $Q$ is given at query time,
and the function of the two inputs must be computed through cell
probes. We restrict the preprocessing and query algorithms to be
deterministic.  In general, we consider a problem in conjunction with
a distribution $\Distr$ over $D \times Q$. Note that the distribution
need not (and, in our case, will not) be a product distribution. We
care about the probability the query algorithm is successful under the
distribution $\Distr$ (for a notion of success to be defined shortly).

As mentioned before, we work in the cell-probe model, and let $\ws$ be
the number of bits in a cell. We assume the query's input consists of
at most $\ws$ bits, and that the space bound is at most $2^{\ws}$. For
the sake of an inductive argument, we extend the cell-probe model by
allowing the data structure to publish some bits at preprocessing
time. These are bits depending on the data structure's input, which
the query algorithm can inspect at no charge.  Closely related to this
concept is our model for a query being successful. We allow the query
algorithm not to return the correct answer, but only in the following
very limited way. After inspecting the query and the published bits,
the algorithm can declare that it cannot answer the query (we say it
{\em rejects} the query). Otherwise, the algorithm can make cell
probes, and at the end it must answer the query correctly. Thus, we
require an a priori admission of any ``error''. In contrast to models
of silent error, it actually makes sense to talk about tiny (close to
zero) probabilities of success, even for problems with boolean output.

For an arbitrary problem $f$ and an integer $k \le 2^{\ws}$, we define
a direct-sum problem $\dsum{k}{f}: D^k \times ([k] \times Q) \to \{ 0,
1 \}$ as follows. The data structure receives a vector of inputs
$(d^1, \dots, d^k)$. The representation depends arbitrarily on all of
these inputs. The query is the index of a subproblem $i \in [k]$, and
an element $q \in Q$. The output of $\dsum{k}{f}$ is $f(q, d^i)$.  We
also define a distribution $\dsum{k}{\Distr}$ for $\dsum{k}{f}$, given
a distribution $\Distr$ for $f$. Each $d^i$ is chosen independently at
random from the marginal distribution on $D$ induced by $\Distr$. The
subproblem $i$ is chosen uniformly from $[k]$, and $q$ is chosen from
the distribution on $Q$ conditioned on $d^i$.

Given an arbitrary problem $f$ and an integer $h \le \ws$, we can
define another problem $\fold{f}{h}$ as follows. The query is a vector
$(q_1, \dots, q_h)$. The data structure receives a regular input $d
\in D$, and integer $r \in [h]$ and the prefix of the query $q_1,
\dots, q_{r-1}$. The output of $\fold{f}{h}$ is $f(d, q_r)$. Note that
we have shared information between the data structure and the querier
(i.e.~the prefix of the query), so $\fold{f}{h}$ is a partial function
on the domain $D \times \, \bigcup_{i=0}^{t-1} Q^i \, \times Q$. Now
we define an input distribution $\fold{\Distr}{h}$ for $\fold{f}{h}$,
given an input distribution $\Distr$ for $f$. The value $r$ is chosen
uniformly at random. Each query coordinate $q_i$ is chosen
independently at random from the marginal distribution on $Q$ induced
by $\Distr$. Now $d$ is chosen from the distribution on $D$,
conditioned on $q_r$.

We give the $\fold{f}{h}$ operator precedence over the direct sum
operator, i.e.~$\dsum{k}{\fold{f}{h}}$ means $\dsum{k}{\left[
\fold{f}{h} \right]}$. Using this notation, we are ready to state our
central cell-probe elimination lemma:

\begin{lemma}  \label{lem:elim}
There exists a universal constant $C$, such that for any problem $f$,
distribution $\Distr$, and positive integers $h$ and $k$, the
following holds. Assume there exists a solution to
$\dsum{k}{\fold{f}{h}}$ with success probability $\delta$ over
$\dsum{k}{\fold{\Distr}{h}}$, which uses at most $k\sigma$ words of
space, $\frac{1}{C} (\frac{\delta}{h})^3 k$ published bits and $T$
cell probes. Then, there exists a solution to $\dsum{k}{f}$ with
success probability $\frac{\delta}{4h}$ over $\dsum{k}{\Distr}$, which
uses the same space, $k \sqrt[h]{\sigma} \cdot C \ws^2$ published bits
and $T-1$ cell probes.
\end{lemma}

\subsection{Setup for the Predecessor Problem}

Let $P(n, \kl)$ be the colored predecessor problem on $n$ integers of
$\kl$ bits each. Remember that this is the decision version of
predecessor search, where elements are colored red or blue, and a
query just returns the color of the predecessor. We first show how to
identify the structure of $\fold{P(n, \kl)}{h}$ inside $P(n, h\kl)$,
making it possible to apply our cell-probe elimination lemma.

\begin{lemma} \label{lem:predfold}
For any integers $n, \kl, h \ge 1$ and distribution $\Distr$ for $P(n,
\kl)$, there exists a distribution $\Distr^{*(h)}$ for $P(n, h\kl)$
such that the following holds. Given a solution to $\dsum{k}{P(n,
h\kl)}$ with success probability $\delta$ over $\dsum{k}{
\Distr^{*(h)}}$, one can obtain a solution to $\dsum{k}{\fold{P(n,
\kl)}{h}}$ with success probability $\delta$ over $\dsum{k}{
\fold{\Distr}{h}}$, which has the same complexity in terms of space,
published bits, and cell probes.
\end{lemma}

\begin{proof}
We give a reduction from $\fold{P(n, \kl)}{h}$ to $P(n, h\kl)$, which
naturally defines the distribution $\Distr^{*(h)}$ in terms of
$\fold{\Distr}{h}$. A query for $\fold{P(n, \kl)}{h}$ consists of
$x_1, \dots, x_h \in \{ 0, 1 \}^\kl$. Concatenating these, we obtain a
query for $P(n, h\kl)$. In the case of $\fold{P(n, \kl)}{h}$, the data
structure receives $i \in [h]$, the query prefix $x_1, \dots, x_{i-1}$
and a set $Y$ of $\kl$-bit integers. We prepend the query prefix to
all integers in $Y$, and append zeros up to $h\kl$ bits. Then, finding
the predecessor of $x_i$ in $Y$ is equivalent to finding the
predecessor of the concatenation of $x_1, \dots, x_h$ in this new set.
\end{proof}

Observe that to apply the cell-probe elimination lemma, the number of
published bits must be just a fraction of $k$, but applying the lemma
increases the published bits significantly. We want to repeatedly
eliminate cell probes, so we need to amplify the number of subproblems
each time, making the new number of published bits insignificant
compared to the new $k$.

\begin{lemma}  \label{lem:probamp}
For any integers $t, \kl, n \ge 1$ and distribution $\Distr$ for $P(n,
\kl)$, there exists a distribution $\Distr^{*t}$ for $P(n\cdot t, \kl
+ \lg t)$ such that the following holds. Given a solution to
$\dsum{k}{P(n\cdot t, \kl + \lg t)}$ with success probability
$\delta$ over $\dsum{k}{\Distr^{*t}}$, one can construct a solution to
$\dsum{kt}{P(n, \kl)}$ with success probability $\delta$ over
$\dsum{kt}{\Distr}$, which has the same complexity in terms of space,
published bits, and cell probes.
\end{lemma}

\begin{proof}
We first describe the distribution $\Distr^{*t}$. We draw $Y_1, \dots,
Y_t$ independently from $\Distr$, where $Y_i$ is a set of integers,
representing the data structures input. Prefix all numbers in $Y_j$ by
$j$ using $\lg t$ bits, and take the union of all these sets to form
the data structure's input for $P(nt, \kl + \lg t)$. To obtain the
query, pick $j \in \{ 0, \dots, t-1 \}$ uniformly at random, pick the
query from $\Distr$ conditioned on $Y_j$, and prefix this query by
$j$. Now note that $\dsum{kt}{\Distr}$ and $\dsum{k}{\Distr^{*t}}$ are
really the same distribution, except that the lower $\lg t$ bits of
the problems index for $\dsum{kt}{\Distr}$ are interpreted as a prefix
in $\dsum{k}{\Distr^{*t}}$. Thus, obtaining the new solution is simply
a syntactic transformation.
\end{proof}

Our goal is to eliminate all cell probes, and then reach a
contradiction. For this, we need the following impossibility result
for a solution making zero cell probes:

\begin{lemma} \label{lem:noprobe}
For any $n \ge 1$ and $\kl \ge \log_2 (n+1)$, there exists a
distribution $\Distr$ for $P(n, \kl)$ such that the following
holds. For all $(\forall) 0 < \delta \le 1$ and $k \ge 1$, there does
not exist a solution to $\dsum{k}{P(n, \kl)}$ with success probability
$\delta$ over $\dsum{k}{\Distr}$, which uses no cell probes and less
than $\delta k$ published bits.
\end{lemma}

\begin{proof}
The distribution $\Distr$ is quite simple: the integers in the set are
always $0$ up to $n-1$, and the query is $n$. All that matters is the
color of $n-1$, which is chosen uniformly at random among red and
blue. Note that for $\dsum{k}{P(n, \kl)}$ there are only $k$ possible
queries, i.e.~only the index of the subproblem matters.

Let $\randp$ be the random variable denoting the published bits. Since
there are no cell probes, the answers to the queries are a function of
$\randp$ alone. Let $\delta(p)$ be the fraction of subproblems that
the query algorithm doesn't reject when seeing the published bits
$p$. In our model, the answer must be correct for all these
subproblems. Then, $\Pr[\randp = p] \le 2^{-\delta(p) k}$, as only
inputs which agree with the $\delta(p) k$ answers of the algorithm can
lead to these published bits. Now observe that $\delta = \E_p
[\delta(p)] \le \E_p \left[ \frac{1}{k} \log_2 \frac{1}{\Pr[\randp =
p]} \right] = \frac{1}{k} H(\randp)$, where $H(\cdot)$ denotes binary
entropy. Since the entropy of the published bits is bounded by their
number (less than $\delta k$), we have a contradiction.
\end{proof}

\subsection{Showing Predecessor Lower Bounds}   \label{sec:pred-proof}

Our proof starts assuming that we for any possible distribution have a
solution to $P(n, \kl)$ which uses $n \cdot 2^a$ space, no published
bits, and successfully answers all queries in $T$ probes, where $T$ is
small. We will then try to apply $T$ {\em rounds\/} of the cell-probe
elimination from Lemma \ref{lem:elim} and \ref{lem:predfold} followed
by the problem amplification from Lemma \ref{lem:probamp}. After $T$
rounds, we will be left with a non-trivial problem but no cell probes,
and then we will reach a contradiction with Lemma \ref{lem:noprobe}.
Below, we first run this strategy ignoring details about the
distribution, but analyzing the parameters for each round. Later in
Lemma \ref{lem:induction}, we will present a formal inductive proof
using these parameters in reverse order, deriving difficult
distributions for more and more cell probes.
 
We denote the problem parameters after $i$ rounds by a subscript
$i$. We have the key length $\kl_i$ and the number of subproblems
$k_i$. The total number of keys remains $n$, so the have $n/k_i$ keys
in each subproblem.  Thus, the problem we deal with in round $i+1$ is
$\dsum{k_i}{P( \frac{n}{k_i}, \kl_i)}$, and we will have some target
success probability $\delta_i$. The number of cells per subproblem is
$\sigma_i= \frac{n}{k_i} 2^a$.  We start the first round with $\kl_0 =
\kl, \delta_0 = 1, k_0 = 1$ and $\sigma_0 = n \cdot 2^a$.

For the cell probe elimination in Lemma \ref{lem:elim} and
\ref{lem:predfold}, our proof will use the same value of $h \ge 2$ in
all rounds. Then $\delta_{i+1} \ge \frac{\delta_i}{4h}$, so $\delta_i
\ge (4h)^{-i}$.  To analyze the evolution of $\kl_i$ and $k_i$, we let
$t_i$ be the factor by which we increase the number of subproblems in
round $i$ when applying the problem amplification from
Lemma \ref{lem:probamp}. We now have $k_{i+1} = t_i \cdot k_i$ and
$\kl_{i+1} = \frac{\kl_i}{h} - \lg t_i$.

When we start the first round, we have no published bits, but when we
apply Lemma \ref{lem:elim} in round $i+1$, it leaves us with up to
$k_i \sqrt[h]{\sigma_i} \cdot Cw^2$ published bits for round $i+2$. We
have to choose $t_i$ large enough to guarantee that this number of
published bits is small enough compared to the number of subproblems
in round $i+2$.  To apply Lemma \ref{lem:elim} in round $i+2$, the
number of published bits must be at most $\frac{1}{C}
(\frac{\delta_{i+1}}{h})^3 k_{i+1} = \frac{\delta_i^3 t_i}{64C h^6}
k_i$.  Hence we must set $t_i \ge \sqrt[h]{\sigma_i} \cdot 64 C^2
\ws^2 h^6 (\frac{1}{\delta_i})^3$. Assume for now that $T = O(\lg
\kl)$. Using $h \le \kl$, and $\delta_i \ge (4h)^{-T} \ge 2^{O(\lg^2
\kl)}$, we conclude it is enough to set:
\begin{equation} \label{eq:cond-ti}
  (\forall) i: \qquad  t_i \ge 
\sqrt[h]{\frac{n}{k_i}} \cdot 2^{a/h} \cdot \ws^2 \cdot 2^{\Theta(\lg^2 \kl)}
\end{equation}

Now we discuss the conclusion reached at the end of the $T$ rounds. We
intend to apply Lemma \ref{lem:noprobe} to deduce that the algorithm
after $T$ stages cannot make zero cell probes, implying that the
original algorithm had to make more than $T$ probes.  Above we made
sure that we after $T$ rounds had $\frac{1}{C} ( \frac{\delta_T}
{h})^3 k_{T} < \delta_T k_T$ published bits, which are few enough
compared to the number $k_T$ of subproblems. The remaining
conditions of Lemma \ref{lem:noprobe} are:
\begin{equation} \label{eq:cond-wn}
  \kl_T \ge 1 \qquad \textrm{and} \qquad \frac{n}{k_T} \ge 1
\end{equation}
Since $\kl_{i+1} \le \frac{\kl_i}{2}$, this condition entails $T =
O(\lg \kl)$, as assumed earlier.

\begin{lemma}  \label{lem:induction} 
With the above parameters satisfying \req{cond-ti} and \req{cond-wn},
for $i = 0,\dots,T$, there is a distribution $\Distr_i$ for $P(
\frac{n}{k_i}, \kl_i)$ so that no solution for $\dsum{k_i}{P(
\frac{n}{k_i}, \kl_i)}$ can have success probability $\delta_i$ over
$\dsum{k_i}{\Distr_i}$ using $n\cdot 2^a$ space, $\frac{1}{C}
(\frac{\delta_{i}}{h})^3 k_{i}$ published bits, and $T-i$ cell probes.
\end{lemma} 

\begin{proof} 
The proof is by induction over $T-i$. A distribution that defies a
good solution as in the lemma is called difficult.  In the base case
$i=T$, the space doesn't matter, and we get the difficult distribution
directly from \req{cond-wn} and Lemma \ref{lem:noprobe}.  Inductively,
we use a difficult distribution $\Distr_i$ to construct a difficult
distribution $\Distr_{i-1}$.

Recall that $k_i=k_{i-1}t_{i-1}$. Given our difficult distribution
$\Distr_i$, we use the problem amplification in Lemma
\ref{lem:probamp}, to construct a distribution $\Distr_i^{*t_{i-1}}$
for $P( \frac{n}{k_i} \cdot t_{i-1}, \kl_i+\lg t_{i-1})=P(
\frac{n}{k_{i-1}}, \kl_i+\lg t_{i-1})$ so that no solution for
$\dsum{k_{i-1}}{P( \frac{n}{k_{i-1}}, \kl_i+\lg t_{i-1})}$ can have
success probability $\delta_i$ over $\dsum{k_{i-1}}{\Distr_i^
{*t_{i-1}}}$ using $n \cdot 2^a$ space, $\frac{1}{C}
(\frac{\delta_i}{h})^3 k_i$ published bits, and $T-i$ cell probes.

Recall that \req{cond-ti} implies $k_{i-1} \sqrt[h]{\sigma_{i-1}}
\cdot Cw^2\leq \frac{1}{C} (\frac{\delta_{i}}{h})^3 k_i$, hence that
$k_{i-1} \sqrt[h]{\sigma_{i-1}}$ is less than the number of bits
allowed published for our difficult distribution $\Distr_i^
{*t_{i-1}}$. Also, recall that $\sigma_j k_j = n\cdot 2^a$ for all
$j$.  We can therefore use the cell probe elimination in Lemma
\ref{lem:elim}, to construct a distribution $\left(\Distr_i^
{*t_{i-1}}\right)^{(h)}$ for $P( \frac{n}{k_{i-1}}, \kl_i+\lg
t_{i-1})^{(h)}$ so that no solution for $\dsum{k_{i-1}}{P(
\frac{n}{k_{i-1}}, \kl_i+\lg t_{i-1})^{(h)}}$ can have success
probability $\delta_{i-1}\geq h\delta_i$ over $\dsum{k_{i-1}}
{\left(\Distr_i^{*t_{i-1}}\right)^{(h)}}$ using $n \cdot 2^a$ space,
$\frac{1}{C} (\frac{\delta_{i-1}}{h})^3 k_{i-1}$ published bits, and
$T-i+1$ cell probes. Finally, using Lemma \ref{lem:predfold}, we use
$\left(\Distr_i^{*t_{i-1}}\right)^{(h)}$ to construct the desired
difficult distribution $\Distr_{i-1}$ for $P( \frac{n}{k_{i-1}},
h(\kl_i+\lg t_{i-1}))=P(\frac{n}{k_{i-1}}, \kl_{i-1})$.
\end{proof}

The predecessor lower bound then follows by applying Lemma
\ref{lem:induction} with $i=0$ and the initial parameters $\kl_0 =
\kl, \delta_0 = 1, k_0 = 1$. We conclude that there is a difficult
distribution $\Distr_0$ for $P(n, \kl)$ with no solution getting
success probability $1$ using $n\cdot 2^a$ space, $0$ published bits,
and $T$ cell probes.

\subsection{Calculating the Trade-Offs}

In this section, we show how to choose $h$ and $t_i$ in order to
maximize the lower bound $T$, under the conditions of \req{cond-ti}
and \req{cond-wn}. First, we show a simple bound on a recursion that
shows up repeatedly in our analysis:

\begin{lemma}  \label{lem:recbnd}
Consider the recursion $x_{i+1} \ge \alpha x_i - \gamma$, for $\gamma
\ge 1$. As long as $i \le \log_{1/\alpha} (\frac{x_0}{1 + \gamma /
(1-\alpha)})$, we have $x_i \ge 1$.
\end{lemma}

\begin{proof}
Expanding the recursion, we have $x_i \ge x_0 \alpha^i - \gamma
(\alpha^{i-1} + \dots + 1) = x_0 \alpha^i - \gamma \frac{1 - \alpha^i}
{1 - \alpha}$. For $x_i \ge 1$, we must have $x_0 \alpha^i \ge 1 +
\gamma \frac{1 - \alpha^i}{1 - \alpha}$, which is true if $x_0
\alpha^i \ge 1 + \frac{\gamma}{1 - \alpha}$. This gives $i \le
\log_{1/\alpha} (\frac{x_0}{1 + \gamma / (1-\alpha)})$.
\end{proof}

We now argue that the bound for low space that we are trying to prove
can only be better than the communication complexity lower bound when
$\lg \kl = O((\lg\lg n)^2)$. This is relevant because our cell-probe
elimination lemma is less than perfect in its technical details, and
cannot always achieve the optimal bound. Fortunately, however, it does
imply an optimal bound when $\kl$ is not too large, and in the
remaining cases an optimal lower bound follows from communication
complexity.

Remember that for space $O(n^2)$, communication complexity implies an
asymptotic lower bound of $\min \{ \frac{\lg n}{\lg \ws},~
\frac{\lg(\kl/\lg n)}{\lg\lg(\kl/\lg n)} \}$. If $\lg \kl =
\Omega((\lg\lg n)^2)$, this is $\Theta( \min \{ \frac{\lg n}{\lg \ws},
\frac{\lg \kl}{\lg\lg \kl} \})$. For $a \le \lg n$, we are trying to
prove an asymptotic lower bound of $\min \{ \frac{\lg n}{\lg \ws},
\frac{\lg(\kl/a)}{\lg(\lg \frac{\kl}{a} / \lg \frac{\lg n}{a})}
\}$. If $\lg \kl = \Omega((\lg\lg n)^2)$, this becomes $\Theta(\min \{
\frac{\lg n}{\lg \ws}, \frac{\lg \kl}{\lg\lg \kl} \})$, which is
identical to the communication bound.

\paragraph{Polynomial Universes.}
Assume that $\kl \ge 3 \lg n$. We first show a lower bound of
$\Omega(\lg \frac{\lg n}{a})$, which matches van Emde Boas on
polynomial universes.  For this, it suffices to set $h = 2$ and $t_i =
(\frac{n}{k_i})^{3/4}$. Then, $\frac{n}{k_{i+1}} =
(\frac{n}{k_i})^{1/4}$, so $\lg \frac{n}{k_i} = 4^{-i} \lg n$ and $\lg
t_i = \frac{3}{4} 4^{-i} \lg n$. By our recursion for $\kl_i$, we have
$\kl_{i+1} = \frac{\kl_i}{2} - \frac{3}{4} 4^{-i} \lg n$. Given $\kl_0
= \kl \ge 3\lg n$, it can be seen by induction that $\kl_i \ge 3 \cdot
4^{-i} \lg n$. Indeed, $\kl_{i+1} \ge 3 \cdot 4^{-i} \cdot \frac{1}{2}
\lg n - \frac{3}{4} 4^{-i} \lg n \ge 3 \cdot 4^{-(i+1)} \lg n$. By the
above, \req{cond-wn} is satisfied for $T \le \Theta(\lg\lg
n)$. Finally, note that condition \req{cond-ti} is equivalent to:
\begin{eqnarray*}
\lg t_i &\ge&
  \frac{1}{h} \lg \frac{n}{k_i} + \frac{a}{h} + \Theta(\lg \ws + \lg^2 \kl)
\quad\Leftrightarrow\quad
  \frac{3}{4} 4^{-i} \lg n ~\ge~ 
  \frac{1}{2} 4^{-i} \lg n + \frac{a}{2} + \Theta(\lg \ws + \lg^2 \kl)
\\ & \Leftrightarrow &
  T ~\le~ \Theta\left(\lg~ \min \left\{ \frac{\lg n}{a}, 
  \frac{\lg n}{\lg^2 w} \right\}\right) =
  \Theta\left( \min \left\{ \lg \frac{\lg n}{a}, \lg\lg n \right\} \right) =
  \Theta\left( \lg \frac{\lg n}{a} \right)
\end{eqnarray*}
Here we have used $\lg w = O((\lg\lg n)^2)$, which is the regime in
which our bound for small space can be an improvement over the
communication bound.

\paragraph{Handling Larger Universes.}
We now show how one can take advantage of a higher $w$ to obtain
larger lower bounds. We continue to assume $w \ge 3\lg n$. Our
strategy is to use the smallest $t_i$ possible according to
\req{cond-ti} and superconstant $h$. To analyze the recursion for
$\kl_i$, we just bound $t_i \le n$, so $\kl_{i+1} \ge \frac{\kl_i}{h}
- \lg n$. Using Lemma \ref{lem:recbnd}, we have $\kl_T \ge 1$ for $T
\le \Theta(\lg_h (\frac{w}{\lg n}))$. We also have the recursion:
\begin{equation*}
\lg \frac{n}{k_{i+1}} = \lg \frac{n}{k_i} - \lg t_i 
= \left( 1 - \frac{1}{h} \right) \lg \frac{n}{k_i} - \frac{a}{h} - O(\lg^2 w)
\end{equation*}
Again by Lemma \ref{lem:recbnd}, we see that $\frac{n}{k_T} \ge 1$ if:
\begin{equation*}
T \le \Theta\left( \lg \frac{\lg n}{h\cdot (\frac{a}{h} + \lg^2 w)} 
  \textrm{\huge ~/}  \lg \frac{1}{1 - \frac{1}{h}} \right)
= \Theta\left( h \lg \frac{\lg n}{a + h\lg^2 w} \right)
= \Theta\left( \min \left\{ h \lg \frac{\lg n}{a}, 
    h\lg \frac{\lg n}{h\lg^2 w} \right\} \right)
\end{equation*}
As mentioned before, the condition $\kl_T \ge 1$ in \req{cond-wn}
implies $T = O(\lg w)$, so we can assume $h = O(\lg w)$. Remember that
we are assuming $\lg w = O((\lg\lg n)^2)$, so the second term in the
min is just $\Theta(h\lg\lg n)$. Then, the entire expression
simplifies to $\Theta(h \lg\frac{\lg n}{a})$.

The lower bound we obtain is be the minimum of the bounds derived by
considering $\kl_i$ and $k_i$. We then choose $h$ to maximize this
minimum, arriving at:
\begin{equation*}
\Theta\left( \max_h \min \left\{ 
   \frac{\lg (w/\lg n)}{\lg h}, h \lg \frac{\lg n}{a} \right\} \right)
\end{equation*}

Clearly, the $\Omega(\lg \frac{\lg n}{a})$ bound derived previously
still holds. Then, we can claim a lower bound that is the maximum of
this and our new bound, or, equivalently up to constants, their sum:
\begin{eqnarray*}
& & \lg\frac{\lg n}{a} + \max_h \min \left\{ \frac{\lg (w/\lg n)}{\lg h},
  h \lg \frac{\lg n}{a} \right\}
= \max_h \min \left\{ \lg\frac{\lg n}{a} + \frac{\lg (w/\lg n)}{\lg h},
  (h+1) \lg \frac{\lg n}{a} \right\} \\
&\ge& \max_h \min \left\{\frac{\lg (w/\lg n) + \lg(\lg n / a)}{\lg h},
  h \lg \frac{\lg n}{a} \right\}
= \max_h \min \left\{\frac{\lg (w/a)}{\lg h},
  h \lg \frac{\lg n}{a} \right\}
\end{eqnarray*}
We choose $h$ to balance the two terms, so $h\lg h = \frac{\lg
(w/a)}{\lg (\lg n / a)}$ and $\lg h = \Theta(\lg\lg \frac{w}{a} -
\lg\lg \frac{\lg n}{a})$. Then the bound is $\Omega(\frac{\lg
(w/a)}{\lg\lg (w/a) - \lg\lg(\lg n / a)})$.

\paragraph{Handling Smaller Universes.}
Finally, we consider smaller universes, i.e.~$w < 3\lg n$. Let $w =
\delta + \lg n$. We start by applying Lemma \ref{lem:probamp} once,
with $t = n / 2^{\delta / 2}$. Now we are looking at the problem
$\dsum{t}{P(2^{\delta / 2}, \frac{3}{2} \delta)}$. Observe that the
subproblems have a universe which is cubic in the number of integers
in the subproblem. Then, we can just apply our strategy for polynomial
universes, starting with $\kl_0 = \frac{3}{2} \delta$ and $n_0 =
2^{\delta / 2}$. We obtain a lower bound of $\Omega(\lg
\frac{\delta}{a}) = \Omega(\lg \frac{w - \lg n}{a})$.

\section{Proof of Cell-Probe Elimination}   \label{sec:cpe-proof}

We assume a solution to $\dsum{k}{\fold{f}{h}}$, and use it to
construct a solution to $\dsum{k}{f}$. The new solution uses the query
algorithm of the old solution, but skips the first cell probe made by
this algorithm. A central component of our construction is a
structural property about any query algorithm for
$\dsum{k}{\fold{f}{h}}$ with the input distribution
$\dsum{k}{\fold{\Distr}{h}}$. We now define and claim this
property. Section \ref{sec:pf-small} uses it to construct a solution
for $\dsum{k}{f}$, while Section \ref{sec:pf-big} gives the proof.

We first introduce some convenient notation. Remember that the data
structure's input for $\dsum{k}{\fold{f}{h}}$ consists of a vector
$(d^1, \dots, d^k) \in D^k$, a vector selecting the interesting
segments $(r^1, \dots, r^k) \in [h]^k$ and the query prefixes $Q^i_j$
for all $j \in [r^i - 1]$. Denote by $\randd, \randr$ and $\randQ$ the
random variables giving these three components of the input. Also let
$\randp$ be the random variable representing the bits published by the
data structure. Note that $\randp$ can also be understood as a
function $\randp(\randd, \randr, \randQ)$. The query consists of an
index $i$ selecting the interesting subproblem, and a vector $(q_1,
\dots, q_h)$ with a query to that subproblem. Denote by $\randi$ and
$\randq$ these random variables.  Note that in our probability space
$\dsum{k}{\fold{f}{h}}$, we have $\randq_j = \randQ^\randi_j,
(\forall) j < \randr^\randi$.

Fix some instance $p$ of the published bits and a subproblem index $i
\in [k]$. Consider a prefix $(q_1, \dots, q_j)$ for a query to this
subproblem. Depending on $q_{j+1}, \dots, q_h$, the query algorithm
might begin by probing different cells, or might reject the query. Let
$\cells^i(p; q_1, \dots, q_j)$ be the set of cells that could be
inspected by the first cell probe. Note that this set could be
$\emptyset$, if all queries are rejected.

Now define:

\begin{equation}  \label{eq:def-epsi}
\eps^i(p) = \left\{
  \begin{array}{ll}
    0 & \textrm{if } \cells^i(p; \randQ^i) = \emptyset \\
    \Pr \left[|\cells^i(p; \randq_1, \dots, \randq_{\randr^i})| \ge
      \frac{\min \{ \sigma, |\cells^i(p; \randQ^i)| \}}{\sqrt[h]{\sigma}} 
      \mid \randi = i \right] & \textrm{otherwise}
  \end{array}
\right.
\end{equation}

The probability space is that defined by $\dsum{k}{\fold{\Distr}{h}}$
when the query is to subproblem $i$. In particular, such a query will
satisfy $\randq_j = \randQ^i_j, (\forall) j < \randr^i$, because the
prefix is known to the data structure. Note that this definition
completely ignores the suffix $\randq_{\randr^i+1}, \dots, \randq_h$
of the query. The intuition behind this is that for any choice of the
suffix, the correct answer to the query is the same, so this suffix can
be ``manufactured'' at will. Indeed, an arbitrary choice of the suffix
is buried in the definition of $\cells^i$.

With these observations, it is easier to understand \req{def-epsi}. If
the data structure knows that no query to subproblem $i$ will be
successful, $\eps_i = 0$. Otherwise, we compare two sets of cells. The
first contains the cells that the querier might probe given what the
data structure knows: $\cells^i(p, \randQ^i)$ contains all cells that
could be probed for various $\randq^i_{\randr^i}$ and various
suffixes. The second contains the cells that the querier could choose
to probe considering its given input $\randq^i_{\randr^i}$ (the
querier is only free to choose the suffix).  Obviously, the second set
is a subset of the first. The good case, whose probability is measured
by $\eps_i$, is when it is a rather large subset, or at least large
compared to $\sigma$.

For convenience, we define $\eps^*(p) = \E_{i \gets [k]}[\eps^i(p)] =
\frac{1}{k} \sum_i \eps^i(p)$. Using standard notation from
probability theory, we write $\eps^i(p \mid E)$, when we condition on
some event $E$ in the probability of \req{def-epsi}. We also write
$\eps^i(p \mid X)$ when we condition on some random variable $X$,
i.e.~$\eps^i(p \mid X)$ is a function $x \mapsto \eps^i(p \mid
X=x)$. We are now ready to state our claim, to be proven in Section
\ref{sec:pf-big}.

\begin{lemma} \label{lem:pf-big}
There exist $\fixr$ and $\fixQ$, such that $\E_{\randd}[
\eps^*(\randp(\fixr, \fixQ, \randd) \mid \randr = \fixr, \randQ =
\fixQ, \randd)] \ge \frac{\delta}{2h}$.
\end{lemma}

\subsection{The Solution for $\dsum{k}{f}$}  \label{sec:pf-small}

As mentioned before, we use the solution for $\dsum{k}{\fold{f}{h}}$,
and try to skip the first cell probe. To use this strategy, we need to
extend an instance of $\dsum{k}{f}$ to an instance of
$\dsum{k}{\fold{f}{h}}$. This is done using the $\fixr$ and $\fixQ$
values whose existence is guaranteed by Lemma \ref{lem:pf-big}. The
extended data structure's input consists of the vector $(d^1, \dots,
d^k)$ given to $\dsum{k}{f}$, and the vectors $\fixr$ and $\fixQ$. A
query's input for $\dsum{k}{f}$ is a problem index $i \in [k]$ and a
$q \in Q$. We extend this to $(q_1, \dots, q_h)$ by letting $q_j =
\fixQ^i_j, (\forall) j < \fixr^i$, and $q_{\fixr^i} = q$, and
manufacturing a suffix $q_{\fixr^i+1}, \dots, q_h$ as described below.

First note that extending an input of $\dsum{k}{f}$ to an input of
$\dsum{k}{\fold{f}{h}}$ by this strategy preserves the desired answer
to a query (in particular, the suffix is irrelevant to the answer).
Also, this transformation is well defined because $\fixr$ and $\fixQ$
are ``constants'', defined by the input distribution
$\dsum{k}{\fold{\Distr}{h}}$.  Since our model is nonuniform, we only
care about the existence of $\fixr$ and $\fixQ$, and not about
computational aspects.

To fully describe a solution to $\dsum{k}{f}$, we must specify how to
obtain the data structure's representation and the published bits, and
how the query algorithm works. The data structure's representation is
identical to the representation for $\dsum{k}{\fold{f}{h}}$, given the
extended input. The published bits for $\dsum{k}{f}$ consist of the
published bits for $\dsum{k}{\fold{f}{h}}$, plus a number of published
cells from the data structure's representation. Which cells are
published will be detailed below. We publish the cell address together
with its contents, so that the query algorithm can tell whether a
particular cell is available.

The query algorithm is now simple to describe. Remember that $q_1,
\dots, q_{\fixr^i - 1}$ are prescribed by $\fixQ$, and $q_{\fixr^i} =
q$ is the original input of $\dsum{k}{f}$. We now iterate through all
possible query suffixes. For each possibility, we simulate the
extended query using the algorithm for $\dsum{k}{\fold{f}{h}}$. If
this algorithm rejects the query, or the first probed cell is not
among the published cells, we continue trying suffixes. Otherwise, we
stop, obtain the value for the first cell probe from the published
cells and continue to simulate this query using actual cell probes. If
we don't find any good suffix, we reject the query. It is essential
that we can recognize success in the old algorithm by looking just at
published bits. Then, searching for a suffix that would not be
rejected is free, as it does not involve any cell probes.

\paragraph{Publishing cells.}
It remains to describe which cells the data structure chooses to
publish, in order to make the query algorithm successful with the
desired probability. Let $p$ be the bits published by the
$\dsum{k}{\fold{f}{h}}$ solution. Note that in order to make the query
$(i,q)$ successful, we must publish one cell from $\cells^i(p;
\fixQ^i, q)$. Here, we slightly abuse notation by letting $\fixQ^i, q$
denote the $\fixr^i$ entries of the prefix $\fixQ^i$, followed by $q$.
We will be able to achieve this for all $(i,q)$ satisfying:
\begin{equation}  \label{eq:when-ok}
  \cells^i(p; \fixQ^i) \ne \emptyset 
  \qquad \textrm{and} \qquad
  |\cells^i(p; \fixQ^i, q)| \ge
    \frac{\min \{ \sigma, |\cells^i(p; \randQ^i)| \}}{\sqrt[h]{\sigma}} 
\end{equation}
Comparing to \req{def-epsi}, this means the success probability is at
least $\eps^*(p \mid \randr = \fixr, \randQ = \fixQ, \randd = (d_1,
\dots, d_k))$. Then on average over possible inputs $(d_1, \dots,
d_k)$ to $\dsum{k}{f}$, the success probability will be at least
$\frac{\delta}{2h}$, as guaranteed by Lemma \ref{lem:pf-big}.

We will need the following standard result:

\begin{lemma}   \label{lem:hitsets}
Consider a universe $U \ne \emptyset$ and a family of sets $\Fam$ such
that $(\forall) S \in \Fam$ we have $S \subset U$ and $|S| \ge
\frac{|U|}{B}$. Then there exists a set $T \subset U, |T| \le B \ln
|\Fam|$ such that $(\forall) S \in \Fam, S \cap T \ne \emptyset$.
\end{lemma}

\begin{proof}
Choose $B \ln |\Fam|$ elements of $U$ with replacement. For a fixed $S
\in \Fam$, an element is outside $S$ with probability at most $1 -
\frac{1}{B}$. The probability all elements are outside $S$ is at most
$(1 - \frac{1}{B})^{B\ln |\Fam|} < e^{-\ln |\Fam|} <
\frac{1}{|\Fam|}$. By the union bound, all sets in $\Fam$ are hit at
least once with positive probability, so a good $T$ exists.
\end{proof}

We distinguish three types of subproblems, parallel to \req{when-ok}.
If $\cells^i(p; \fixQ^i) = \emptyset$, we make no claim (the success
probability can be zero). Otherwise, if $|\cells^i(p; \fixQ^i)| <
\sigma$, we handle subproblem $i$ using a local strategy. Consider all
$q$ such that $|\cells^i(p; \fixQ^i, q)| \ge \frac{|\cells^i(p;
\fixQ^i)|}{\sqrt[h]{\sigma}}$. We now apply Lemma \ref{lem:hitsets}
with the universe $\cells^i(p; \fixQ^i)$ and the family $\cells^i(p;
\fixQ^i, q)$, for all interesting $q$'s. There are at most $2^w$
choices of $q$, bounding the size of the family. Then, the lemma
guarantees that the data structure can publish a set of
$O(\sqrt[h]{\sigma} \cdot w)$ cells which contains at least one cell
from each interesting set. This means that each interesting $q$ can be
handled successfully by the algorithm.

We handle the third type of subproblems, those with $|\cells^i(p;
\fixQ^i)| \ge \sigma$, in a global fashion. Consider all
``interesting'' pairs $(i,q)$ with $|\cells^i(p; \fixQ^i, q)| \ge
\sigma^{1 - 1/h}$. We now apply Lemma \ref{lem:hitsets} with the
universe consisting of all $k\sigma$ cells, and the family being
$\cells^i(p; \fixQ^i, q)$, for interesting $(i,q)$. The cardinality of
the family is at most $2^w$, since $i$ and $q$ form a query, which
takes at most one word. Then by Lemma \ref{lem:hitsets}, the data
structure can publish a set of $O(k \sqrt[h]{\sigma} \cdot w)$ cells,
which contains at least one cell from each interesting set. With these
cells, the algorithm can handle successfully all interesting $(i,q)$
queries.

The total number of cells that we publish is $O(k \sqrt[h]{\sigma}
\cdot w)$. Thus, we publish $O(k \sqrt[h]{\sigma} \cdot w^2)$ new
bits, plus $O(k)$ bits from the assumed solution to
$\dsum{k}{\fold{f}{h}}$. For big enough $C$, this is at most
$k\sqrt[h]{\sigma} \cdot C w^2$.

\subsection{An Analysis of $\dsum{k}{\fold{f}{h}}$: 
  Proof of Lemma \latexref{lem:pf-big}} \label{sec:pf-big}

Our analysis has two parts. First, we ignore the help given by the
published bits, by assuming they are constantly set to some value
$p$. As $\randr^i$ and $\randQ^i$ are chosen randomly, we show that
the conditions of \req{def-epsi} are met with probability at least
$\frac{1}{h}$ times the success probability for subproblem $i$. This
is essentially a lower bound on $\eps^i$, and hence on $\eps^*$.

Secondly, we show that the published bits do not really affect this
lower bound on $\eps^*$. The intuition is that there are two few
published bits (much fewer than $k$) so for most subproblems they are
providing no information at all. That is, the behavior for that
subproblem is statistically close to when the published bits would not
be used. Formally, this takes no more than a (subtle) application of
Chernoff bounds. The gist of the idea is to consider some setting $p$
for the published bits, and all possible inputs (not just those
leading to $p$ being published). In this probability space, $\eps^i$
are independent for different $i$, so the average is close to $\eps^*$
with overwhelmingly high probability. Now pessimistically assume all
inputs where the average of $\eps^i$ is not close to $\eps^*$ are
possible inputs, i.e.~input for which $p$ would be the real help
bits. However, the probability of this event is so small, that even
after a union bound for all $p$, it is still negligible.

We now proceed to the first part of the analysis. Let $\delta^i(p)$ be
the probability that the query algorithm is successful when receiving
a random query for subproblem $i$. Formally, $\delta^i(p) = \Pr [
\cells^i(p; \randq) \ne \emptyset \mid \randi = i]$. We define
$\delta^i(p \mid E), \delta^i(p \mid X)$ and $\delta^*(\cdot)$ similar
to the functions associated to $\eps^i$. Observe that the probability
of correctness guaranteed by assumption is $\delta = \E_{\randr,
\randQ, \randd}[ \delta^*(\randp(\randr, \randQ, \randd) \mid \randr,
\randQ, \randd)]$.

\begin{lemma}  \label{lem:trie}
For any $i$ and $p$, we have $\eps^i(p) \ge \frac{\delta^i(p)}{h}$.
\end{lemma}

\begin{proof}
Let us first recall the random experiment defining $\eps^i(p)$. We
select a uniformly random $r \in [h]$ and random $q_1, \dots,
q_{r-1}$. First we ask whether $\cells^i(p; q_1, \dots, q_{r-1}) =
\emptyset$. If not, we ask about the probability that a random $q_r$
is good, in the sense of \req{def-epsi}. Now let us rephrase the
probability space as follows: first select $q_1, \dots, q_h$ at
random; then select $r \in [h]$ and use just $q_1, \dots, q_r$ as
above. The probability that the query $(q_1, \dots, q_h)$ is handled
successfully is precisely $\delta^i(p)$. Let's assume it
doesn't. Then, for any $r$, $\cells^i(p; q_1, \dots, q_{r-1}) \ne
\emptyset$ because there is at least one suffix which is handled
successfully. We will now show that there is at least one choice of
$r$ such that $q_r$ is good when the prefix is $q_1, \dots,
q_{r-1}$. When averaged over $q_1, \dots, q_{r-1}$, this gives a
probability of at least $\frac{\delta^i(p)}{h}$

To show one good $r$, let $\phi_r = \min \{ |\cells^i(p; q_1, \dots,
q_{r-1})|, \sigma \}$. Now observe that $\frac{\phi_1}{\phi_2} \cdot
\frac{\phi_2}{\phi_3} \cdot \dots \cdot \frac{\phi_{h-1}}{\phi_h} =
\frac{\phi_1}{\phi_h} \le \phi_1 \le \sigma$. By the pigeonhole
principle, $(\exists) r: \frac{\phi_{r}}{\phi_{r+1}} \le \sigma^{1/h}$. 
This implies $|\cells^i(p; q_1, \dots, q_r)| \ge \frac{\min \{ \sigma,
|\cells^i(p; q_1, \dots, q_{r-1})|}{\sqrt[h]{\sigma}}$, as desired.
\end{proof}

Note that if the algorithm uses zero published bits, we are
done. Thus, for the rest of the analysis we may assume $\frac{1}{C}
(\frac{\delta}{h})^3 k \ge 1$.  We now proceed to the second part of
the analysis, showing that $\eps^*$ is close to the lower bound of the
previous lemma, even after a union bound over all possible published
bits.

\begin{lemma}  \label{lem:chernoff1}
With probability at least $1 - \frac{\delta}{8h}$ over random $\randr,
\randQ$ and $\randd$:
$ (\forall) p:~ \eps^*(p \mid \randr, \randQ, \randd) 
  \ge \frac{\delta^*(p)}{h} - \frac{\delta}{4h} $
\end{lemma}

\begin{proof}
Fix $p$ arbitrarily. By definition, $\eps^*(p \mid \randr, \randQ,
\randd) = \frac{1}{k} \sum_i \eps^i(p \mid \randr, \randQ,
\randd)$. By Lemma \ref{lem:trie}, $\E[\eps^i(p \mid \randr, \randQ,
\randd)] = \eps^i(p) \ge \frac{\delta^i(p)}{h}$, which implies
$\eps^*(p) \ge \frac{\delta^*(p)}{h}$. Thus, our condition can be
rephrased as:
\[ \frac{1}{k} \sum_i \eps^i(p \mid \randr, \randQ, \randd)
 \ge \E\left[ \frac{1}{k} \sum_i \eps^i(p \mid \randr, \randQ, \randd)
   \right] - \frac{\delta}{4h}
\]
Now note that $\eps^i(p \mid \randr, \randQ, \randd)$ only depends on
$\randr^i, \randQ^i$ and $\randd^i$, since we are looking at the
behavior of a query to subproblem $i$ for a fixed value of the
published bits; see the definition of $\eps^i$ in \req{def-epsi}.
Since $(\randr^i, \randQ^i, \randd^i)$ are independent for different
$i$, it follows that $\eps^i(p \mid \randr, \randQ, \randd)$ are also
independent. Then we can apply a Chernoff bound to analyze the mean
$\eps^*(p \mid \randr, \randQ, \randd)$ of these independent random
variables. We use an additive Chernoff bound \cite{alon-spencer}:
\[ \Pr_{\randr, \randQ, \randd} \left[
  \eps^*(p \mid \randr, \randQ, \randd) < \eps^*(p) - \frac{\delta}{4h}
 \right] < e^{-\Omega(k (\frac{\delta}{h})^2)} \]
Now we take a union bound over all possible choices $p$ for the
published bits. The probability of the bad event becomes
$2^{\frac{1}{C} (\frac{\delta}{h})^3 k} e^{-\Omega((
\frac{\delta}{h})^2 k)}$. For large enough $C$, this is $\exp(
-\Omega((\frac{\delta}{h})^2 k))$, for any $\delta$ and $h$.  Now we
use that $\frac{1}{C} (\frac{\delta}{h})^3 k \ge 1$, from the
condition that there is at lest one published bit, so this probability
is at most $e^{-\Omega(C h / \delta)}$. Given that $\frac{h}{\delta}
\ge 1$, this is at most $\frac{\delta}{8h}$ for large enough $C$.
\end{proof}

Unfortunately, this lemma is not exactly what we would want, since it
provides a lower bound in terms of $\delta^*(p)$. This probability of
success is measured in the original probability space. As we condition
on $\randr, \randQ$ and $\randd$, the probability space can be quite
different. However, we show next that in fact $\delta^*$ cannot change
too much. As before, the intuition is that there are too few published
bits, so for most subproblems they are not changing the query
distribution significantly.

\begin{lemma}  \label{lem:chernoff2}
With probability at least $1 - \frac{\delta}{8}$ over random $\randr,
\randQ$ and $\randd$: 
$ (\forall) p:~ \delta^*(p \mid \randr, \randQ, \randd) 
\le \delta^*(p) + \frac{\delta}{4} $
\end{lemma}

\begin{proof}
The proof is very similar to that of Lemma \ref{lem:chernoff1}. Fix
$p$ arbitrarily. By definition, $\delta^*(p \mid \randr, \randQ,
\randd)$ is the average of $\delta^i(p \mid \randr, \randQ,
\randd)$. Note that for fixed $p$, $\delta^i$ depends only on
$\randr^i, \randQ^i$ and $\randd^i$. Hence, the $\delta^i$ values are
independent for different $i$, and we can apply a Chernoff bound to
say the mean is close to its expectation. The rest of the calculation
is parallel to that of Lemma \ref{lem:chernoff1}.
\end{proof}

We combine Lemmas \ref{lem:chernoff1} and \ref{lem:chernoff2} by a
union bound. We conclude that with probability at least $1 -
\frac{\delta}{4}$ over random $\randr, \randQ$ and $\randd$, we have
that $(\forall) p$:
\begin{equation}  \label{eq:eps-delta}
  \left. \begin{array}{l}
    \eps^*(p \mid \randr, \randQ, \randd) 
      \ge \frac{\delta^*(p)}{h} - \frac{\delta}{4h} \\[.5ex]
    \delta^*(p \mid \randr, \randQ, \randd) 
      \le \delta^*(p) + \frac{\delta}{4}
  \end{array} \right\}  \Rightarrow
  \eps^*(p \mid \randr, \randQ, \randd) - 
    \frac{\delta^*(p \mid \randr, \randQ, \randd)}{h} \ge -\frac{\delta}{2h}
\end{equation}

Since this holds for all $p$, it also holds for $p = \randp$, i.e.~the
actual bits $\randp(\randr, \randQ, \randd)$ published by the data
structure given its input. Now we want to take the expectation over
$\randr, \randQ$ and $\randd$. Because $\eps^*(\cdot), \delta^*(\cdot)
\in [0,1]$, we have $\eps^*(\cdot) - \frac{1}{h} \delta^*(\cdot) \ge
-\frac{1}{h}$. We use this as a pessimistic estimate for the cases of
$\randr, \randQ$ and $\randd$ where \req{eps-delta} does not hold. We
obtain:
\begin{eqnarray*}
  & & \E\left[ \eps^*(\randp \mid \randr, \randQ, \randd) - 
    \frac{\delta^*(\randp \mid \randr, \randQ, \randd)}{h} \right]
  \ge -\frac{\delta}{2h} + \frac{\delta}{4} \cdot \left( -\frac{1}{h} \right)
  = -\frac{3\delta}{4h} 
  \\ & \Rightarrow &
  \E \big[ \eps^*(\randp \mid \randr, \randQ, \randd) \big]
  \ge \frac{1}{h} \E \big[\delta^*(\randp \mid \randr, \randQ, \randd) \big]
    - \frac{3\delta}{4h} 
  = \frac{1}{h} \delta - \frac{3\delta}{4h} 
  = \frac{\delta}{4h}
\end{eqnarray*}

\section{Communication Lower Bounds}

\subsection{Protocol Manipulations}

To obtain our improved lower bounds for large space, we use two-party
communication complexity. In this section, we state the protocol
manipulation tools that we will use in our proof. We allow protocols
to make errors, and look at the error probability under appropriate
input distributions. Thus, as opposed to our lower bounds for small
space, we also obtain lower bounds for randomized algorithms with
bounded error. We define an $[A; m_1, m_2, m_3, \dots]$-protocol to be
a protocol in which Alice speaks first, sending $m_1$ bits, Bob then
sends $m_2$ bits, Alice sends $m_3$ bits and so on. In a $[B; m_1,
m_2, \dots]$-protocol, Bob begins by sending $m_1$ bits.

For a communication problem $f : A \times B \to \{ 0, 1 \}$, define a
new problem $f^{A,(k)}$ in which Alice receives $x_1, \dots, x_k \in
A$, Bob receives $y \in B, i \in [k]$ and $x_1, \dots, x_{i-1}$, and
they wish to compute $f(x_i, y)$. This is similar to our definition
for $f^{(t)}$, except that we need to specify that Alice's input is
being multiplied. We define $f^{B,(k)}$ symmetrically, with the roles
of Alice and Bob reversed. Finally, given a distribution $\Distr$ for
$f$, we define $\Distr^{A,(k)}$ and $\Distr^{B,(k)}$ following our old
definition for $\Distr^{(k)}$.

The first tool we use is round elimination, which, as mentioned
before, has traditionally been motivated by predecessor lower
bounds. The following is a strong version of this result, due to
\cite{sen-roundelim}:

\begin{lemma}[round elimination \cite{sen-roundelim}]
Suppose $f^{(k),A}$ has an $[A; m_1, m_2, \dots]$-protocol with error
probability at most $\eps$ on $\Distr^{A,(k)}$. Then $f$ has a $[B;
m_2, \dots]$-protocol with error probability at most $\eps +
O(\sqrt{\frac{m_1}{k}})$ on $\Distr$.
\end{lemma}

As opposed to previous proofs, we also bring message compression into
play. The following is from \cite{chakrabarti04ann}, restated in terms
of our $f^{A,(k)}$ problem:

\begin{lemma}[message compression \cite{chakrabarti04ann}]
Suppose $f^{(k),A}$ has an $[A; m_1, m_2, \dots]$ protocol with error
probability at most $\eps$ on $\Distr^{A,(k)}$. Then for any $\delta >
0$, $f$ has an $[A; O(\frac{1 + (m_1 / k)}{\delta^2}), m_2,
\dots]$-protocol with error probability at most $\eps + \delta$ on
$\Distr$.
\end{lemma}

Since this lemma does not eliminate Alice's message, but merely
reduces it, it is used in conjuction with the message switching
technique \cite{chakrabarti04ann}. If Alice's first message has $a$
bits, we can eliminate it if Bob sends his reply to all possible
messages from Alice (thus increasing his message by a factor of
$2^a$), and then Alice includes her first message along with the
second one (increasing the second message size additively by $a$):

\begin{lemma}[message switching]
Suppose $f$ has an $[A; m_1, m_2, m_3, m_4, \dots]$-protocol. Then it
also has a $[B; 2^{m_1} m_2, m_1 + m_3, m_4, \dots]$-protocol with the
same error complexity.
\end{lemma}

Message compression combined with message switching represent, in some
sense, a generalization of the round elimination lemma, allowing us to
trade a smaller $k$ for a larger penalty in Bob's messages. However,
the trade-off does yield round elimination as the end-point, because
message compression cannot reduce Alice's message below
$\Omega(\delta^{-2})$ for any $k$. We combine these two lemmas to
yield a smooth trade-off (with slightly worse error bounds), which is
easier to work with:

\begin{lemma}   \label{lem:cmp-switch}
Suppose $f^{(k)}$ has an $[A; m_1, m_2, m_3, m_4, \dots]$-protocol
with error $\eps$ on $\Distr^{A,(k)}$. Then for any $\delta > 0$, $f$
has a $[B; 2^{O(m_1 / (k\delta^4))} m_2, m_1 + m_3, m_4,
\dots]$-protocol with error probability $\eps + \delta$ on $\Distr$.
\end{lemma}

\begin{proof}
If $\frac{m_1}{k} \le \delta^2$, we can apply the round elimination
lemma. Then, Alice's first message is ommitted with an error increase
of at most $\delta$. None of the subsequent messages change.  If
$\frac{m_1}{k} \ge \delta^2$, we apply the message compression lemma,
which reduces Alice's first message to $O(\frac{1 + (m_1/k)}
{\delta^2})$ bits, while increasing the error by $\delta$. Since
$\frac{m_1}{k} \ge \delta^2$, the bound on Alice's message is at most
$O(\frac{m_1}{k \delta^4})$. Then, we can eliminate Alice's first
message by switching. Note our bound for the second message from Alice
is loose, since it ignores the compression we have done.
\end{proof}

\subsection{Application to Predecessor Search}

\begin{theorem}
Consider a solution to colored predecessor search in a set of $n$
$\kl$-bit integers, which uses space $n\cdot 2^a$ in the cell-probe
model with cells of $\ws$ bits. If $a = \Omega(\lg n)$ and the query
algorithm has an error probability of at most $\frac{1}{3}$, the query
time must satisfy:
\begin{equation*}
T = \Omega \left( \min \left\{ \lg_{\ws} n,~
      \frac{\lg(\kl / a)}{\lg\lg(\kl / a) + \lg(a / \lg n)} \right\} \right)
\end{equation*}
\end{theorem}

\begin{proof}
We consider the communication game in which Alice receives the query
and Bob receives the set of integers. Alice's messages will have $\lg
(n \cdot 2^a) = \Theta(a)$ bits, and Bob's $\ws$ bits. The structure
of our proof is similar to the application of the cell-probe
elimination lemma in Section \ref{sec:smallspace}. By Lemma
\ref{lem:predfold}, we can identify the structure of
$P(n,\kl)^{A,(h)}$ in $P(n, h\kl)$. Then, we can will apply our Lemma
\ref{lem:cmp-switch} to eliminate Alice's messages. Now, we use Lemma
\ref{lem:probamp} to identify the structure of $P(n,\kl)^{B,(t)}$ in
$P(n\cdot t, \kl + \lg t)$. Note that $P(n,\kl)^{B,(t)}$ is
syntactically equivalent to our old $\dsum{t}{P(n,\kl)}$, except that
Alice also receives a (useless) prefix of Bob's input. Now we apply
the round elimination lemma to get rid of Bob's message. 

Thus, after eliminating a message from each player, we are left with
another instance of the colored predecessor problem, with smaller $n$
and $\kl$ parameters. This contrasts with our cell-probe proof, which
couldn't work with just one subproblem, but needed to look at all of
them to analyzing sharing. Our strategy is to increase the error by at
most $\frac{1}{9T}$ in each round of the previous argument. Then,
after $T$ steps, we obtain an error of at most $\frac{1}{3} +
\frac{1}{9} < \frac{1}{2}$. Assuming we still have $n \ge 2$ and $\kl
\ge 1$, it is trivial to make the answer to the query be either red or
blue with equal probability. Then, no protocol with zero communication
can have error complexity below $\frac{1}{2}$, so the original
cell-probe complexity had to be greater than $T$.

As explained in Section \ref{sec:pred-proof}, the proof should be
interpreted as an inductive argument in the reverse direction.
Assuming we have a distribution on which no protocol with $i$ rounds
can have error less than $\eps$, our argument constructs a
distribution on which no protocol with $i+1$ rounds can have error
less than $\eps - \frac{1}{9T}$. At the end, we obtain a distribution
on which no protocol with $T$ rounds can have error $\frac{1}{3}$,
implying the cell-probe lower bound.

It remains to define appropriate values $h$ and $t$ which maximize our
lower bound $T$ by the above discussion. After step $i$, Alice's
message will have size $(i+1) \cdot (a + \lg n) = O(aT)$, because we
have applied message switching $i$ times (in the form of Lemma
\ref{lem:cmp-switch}). Applying Lemma \ref{lem:cmp-switch} one more
time with $\delta = \frac{1}{18T}$, we increases Bob's next message to
$\ws \cdot 2^{O(a T^5 / h)}$.  We now apply round elimination to get
rid of Bob's message. We want an error increase of at most
$\frac{1}{18T}$, adding up to at most $\frac{1}{9T}$ per round. Then,
we set $t$ according to:
\begin{equation*}
O\left( \sqrt{ \frac{\ws \cdot 2^{O(a T^5 / h)}}{t} } \right) 
\le \frac{1}{18T}
  \quad\Rightarrow\quad
t = \ws \cdot O(T^2) \cdot 2^{O(a T^5 / h)}
\end{equation*}

Let $n_i$ and $\kl_i$ denote the problem parameters after $i$ steps of
our argument. Initially, $n_0 = n, \kl_0 = \kl$. By the discussion
above, we have the recursions: $\kl_{i+1} = \frac{\kl_i}{h} - \lg t$
and $\lg n_{i+1} = \lg \frac{n_i}{t} = \lg n_i - \lg t$.

We have $\lg t = O(\lg \ws + \lg T + \frac{aT^5}{h})$. Since we want
$\kl_T \ge 1$, we must have $T \le \lg \kl \le \lg \ws$, so $\lg t$
simplifies to $O(\lg \ws + \frac{aT^5}{h})$. Now the condition $n_T
\ge 2$ implies the following bound on $T$:
\begin{equation*}
T < \frac{\lg n - 1}{\Theta( \lg \ws + \frac{a T^5}{h} )} 
= \Theta \left( \min \left\{ 
       \frac{\lg n}{\lg \ws},\ \frac{h\lg n}{a T^5}
    \right\} \right)
\end{equation*}
To analyze the condition $\kl_T \ge 1$, we apply the recursion bound
of Lemma \ref{lem:recbnd}, implying $T < \lg_h (\frac{\kl}
{\Theta(\lg t)})$. This is satisfied if we upper bound $\lg t$ by
$O((a + \lg \ws) \cdot (1 + \frac{T^5}{h}))$, and set:
\begin{equation*}
T < \Theta\left( \frac{ \lg(\frac{\kl}{a + \lg \ws}) 
  - \lg(\frac{T^5}{h})}{\lg h} \right) =
\Theta\left( \frac{\lg( \frac{\kl}{a + \lg \ws} )}{\lg h} \right) - O(\lg T)
\quad\Rightarrow\quad
T < \Theta\left( \frac{\lg( \frac{\kl}{a + \lg \ws} )}{\lg h} \right)
\end{equation*}

Thus, our lower bound is, up to constant factors, $\min \{ \frac{\lg
n}{\lg \ws},~ \frac{h\lg n}{a T^5},~ \frac{\lg(\kl / (a + \lg
\ws))}{\lg h} \}$. First we argue that we can simplify $a + \lg\ws$ to
just $a$ in the last term. If $a = \Omega(\lg \ws)$, this is trivial.
Otherwise, we have $a = O(\lg \ws)$, so $\lg n = O(\lg \ws)$. But in
this case the first term of the min is $O(1)$ anyway, so the other
terms are irrelevant.

It now remains to choose $h$ in order to maximize the lower bound.
This is achieved when $\frac{h\lg n}{a T^5} = \frac{\lg(\kl / a)}{\lg
h}$, so we should set $\lg h = \Theta(\lg\lg \frac{\kl}{a} +
\lg\frac{a}{\lg n} + \lg T)$. The $\lg T$ term can be ignored because
$T = O(\lg \frac{\kl}{a})$. With this choice of $h$, the lower bound
becomes, up to constants, $\min \{ \frac{\lg n}{\lg \ws}, \frac{
\lg(\kl / a)}{\lg\lg(\kl / a) + \lg(a / \lg n)} \}$.
\end{proof}

\section{Upper Bounds} 

We are working on the static predecessor problem where we are first
given a set $Y$ of $n$ keys. The predecessor of a query key $x$ in $Y$
is the largest key in $Y$ that is smaller than or equal to $x$.  If
$x$ is smaller than any key in $Y$, its predecessor is $-\infty$,
representing a value smaller than any possible key. Below each key is
assumed to be a non-negative $\ell$-bit integer. We are working
on a RAM with word length $\ws\geq\ell$. The results also apply in
the stronger external memory model where $\ws$ is the bit size
of a block. The external memory model is stronger because it like
the cell-probe model does not count computations.

For $n\leq s$ and $\lg n\leq \ell\leq\ws$, we will show represent $n$
$\kl$-bit keys using $O(s\kl)$ bits of space where $s\geq n$.  With $a=\lg
\frac{sw}n$, we will show how to search predecessors in time
\begin{eqnarray}
O\left(\frac {\lg n}{\lg \ws}\right)&&\label{eq:FWB}\\[2ex]
O\left(\lg \frac {\kl-\lg n}a\right)&\textnormal{if}&
\kl=O(\lg n)\label{eq:vEB}\\[2ex]
O\left(\frac{\lg \frac \kl a}{\lg \frac{a\lg \frac \kl a}{(\lg n)}}\right)&
\textnormal{if}&a\geq\lg n\textnormal{ \ and \ }\kl=
\omega(\lg n)\label{eq:large-space}\\[2ex]
O\left(\frac{\lg \frac {\kl}a}{\lg\frac{\lg \frac {\kl}a}
{\lg\frac{\lg n}{a}}}\right)&
\textnormal{if}&a\leq\lg n\textnormal{ \ and \ }\kl=
\omega(\lg n)\label{eq:small-space}
\end{eqnarray}

\paragraph{Contents}
Below, we first obtain \req{FWB} using either B-trees or the fusion
trees of Fredman and Willard \cite{fredman93fusion}. Next we use
\req{FWB} to increase the space by a factor $\ws$ so that we have
$O(2^a\ell)$ bits of space available per key. Then we prove
\req{vEB} by a slight tuning of van Emde Boas' data structure
\cite{vEB77pred}. This bound is tight when $\ws=O(\lg n)$. Next,
elaborating on techniques of Beame and Fich \cite{beame02pred}, we
will first show \req{large-space} and then \req{small-space} in
the case where $\ws\geq2\lg n$.

\subsection{Preliminaries}
In our algorithms, we will assume that $\ws$, $\kl$, and $a$ are
powers of two. For the word length $\ws$, we note we that can simulate
up to twice the word length implementing each extended word operation
with a constant number of regular operations. Hence, internally, our
algorithms can use a word length rounded up to the nearest power of
two without affecting the asymptotic search times. Concerning
the parameters $\kl$ and $a$, we note that it does not affect
the asymptotics if they change by constant factors, so we
can freely round $\kl$ up to the nearest factor of two and $a$ down
to the nearest factor of two, thus accepting a larger key length and
lesser space for the computations.

The search times will be achieved via a series of reductions that often
reduce the key length $\ell$. We will make sure that each reduction
is by a power of two.
  
We will often allocate arrays with $m$ entries,
each of $\ell$ bits. These occupy $m\ell$ consecutive bits in
memory, possibly starting and ending in the middle of words. 
As long as $\ell\leq \ws$, using simple
arithmetic and shifts, we can access or change an entry in constant time.
In our case, the calculations are particularly simple because $\kl$ and
$\ws$ are powers of two.

We will use product notation for concatenation
of bit strings. Hence $xy$ or $x\concat y$ denotes the concatenation of
bit strings $x$ and $y$. As special notation, we
define $-\infty\concat x=x\concat -\infty=-\infty$. 

Finally, we define $\log=\log_2$. Note that this is different from $\lg$ which
is the function used in our asymptotic bounds.

\subsection{Fusion or B-trees}\label{S:FWB}
With Fredman and Willard's fusion trees \cite{fredman93fusion}, we
immediately get a linear space predecessor search time of
$O\left(\frac {\lg n}{\lg \kl}\right)$. If $\ws\leq\kl^2$, this
implies the $O\left(\frac {\lg n}{\lg \ws}\right)$ search time from
\req{FWB}.  Otherwise, we use a B-tree of degree $d=\ws/\kl$. We
can pack the $d$ keys in a singe word, and we can then search a B-tree
node in constant time using some of the simpler bit manipulation from
\cite{fredman93fusion}.  This gives a search time of $O\left(\frac
{\lg n}{\lg d}\right)$ which is $O\left(\frac {\lg n}{\lg\ws}\right)$
for $\ws\geq\kl^2$. Thus we achieve the search time from \req{FWB}
using only linear space, or $O(n\ell)$ bits.

We note that the B-tree solution is simpler in the external memory model
where we do not worry about the actual computations.

\subsection{Inceasing the space}\label{S:space-inc}
We will now use \req{FWB} to increase the space per key by a factor
$\ws$. We simply pick out a set $Y'$ of $n'=\lfloor n/\ws\rfloor$
equally spaced keys so that we have a segment of less than $\ws$ keys
between consecutive keys in $Y'$. We will first do a predecessor
search in $Y'$, and based on the result, do a predecessor search in
the appropriate segment. Since the segment has less than $\ws$ keys,
by \req{FWB}, it can be searched in constant time.

Thus we are left with the problem of doing a predecessor search in
$Y'$.  For this we have $O(s\ell)$ bits, which is $O(s\ell/(n/\ws))=
O(s\ws\ell/n)=O(2^a\ell)$ bits per key. Moreover we note that
replacing $n$ by $n'=\lfloor n/\ws\rfloor$ does not increase any of
the bounds \req{vEB}-\req{small-space}. Hence it suffices
to prove the these bounds \req{vEB}-\req{small-space} assuming 
that we $O(n2^a\ell)$ bits of space available.

\subsection{A tuned van Emde Boas bound for polynomial universes}\label{S:vEB}
In this section, we develop a tuned version of van Emde
Boas's data structure, representing $n$ keys in $O\left(n2^a\ell\right)$ bits
of space providing the search time from \req{vEB} of
\[O\left(\lg\frac{\kl-\lg n}{a}\right).\]
We shall only use this bound for polynomial universes, that is,
when $\kl=O(\lg n)$.

\subsubsection{Complete tabulation}\label{S:tab} The static predecessor
problem is particularly easy when we have room for a complete tabulation
of all possible query keys, that is, if we have
$O(2^\kl\kl)$ bits of space. Then we can allocate a table $\fpred_Y$ that 
for each possible query key $x$ stores the predecessor
$\fpred_Y[x]$ of $x$ in $Y$. If $x<\min Y$, $\fpred_Y[x]=-\infty$.
For our bounds, we will use this as a base 
case if $\kl\leq a$. 

Note that this simple base case is a prime example of what we can do when not
restricted to comparisons on a pointer machine: we use the key as an
addrees to a table entry and get the answer in constant time.

\subsubsection{Prefixes tabulation}\label{S:pref-tab}
If our keys are too long for a complete tabulation, but not too
much longer, it may still be relevant to use tabulation based
on the first $p$ bits of each key. Below it is understood that
the prefix of a key is the first $p$ bits and the suffix is the
last $\kl-p$ bits. Let $\fSuff_Y[u]$ be the suffixes in $Y$ of
keys with prefix $u$. Also let $\fspred_Y[u]$ to denote the strict predessor 
in $Y$ of $u$ suffixed by zeroes. Here by strict predecessor, 
we mean an unequal predecessor. If no length $p$ prefix in $Y$ is smaller 
than $u$,  $\fspred_Y[u]=-\infty$.
The representation of $Y$ now consists of the table that with
each prefix $u$ associates $\fspred_Y[u]$ and 
a  recursive representatation of $\fSuff_Y[u]$. Note that if
$\fSuff_Y[u]=\emptyset$, the recursive representation returns
$-\infty$ on any predecessor query.

We now have the following pseudo-code for searching $Y$:
\begin{quote}
$\fPred(x,Y)$\\
$\null\quad$ $(x_0,x_1)=(\fprefix(x),\fsuffix(x))$\\
$\null\quad$ $y_1=\fPred(x_1,\fSuff_Y[x_0])$\\
$\null\quad$ if $y_1=-\infty$ then return $\fspred_Y[x_0]$.\\
$\null\quad$ return $x_0\concat y_1$\\
\end{quote}
Note in the above pseudo-code that the produre termintes as
soon as it executes return statement. Hence the last statement
is only executed if $y_1\neq-\infty$. Also, as a rule of thumb,
we use square brackets around the argument of a function that we can compute
in constant time..

We shall use this reduction with $p\approx \lg n$ as the first step of our
predecessor search. More precisely, we choose $p\leq\lg n$ such that
the reduced length $\kl-p$ is a power of two less than $2(\kl-\lg n)$.
The reduction adds a constant to the search time. It uses 
$O(2^p\ell)=O(n\ell)$ bits of space on the tables over the prefixes. 
The suffix of a key $y$ appears in the subproblem $\fSuff_Y[u]$. Hence 
the subproblems have a total of $n$ keys, each of length less than $2(\kl-\lg n)$.

\subsubsection{Van Emde Boas' reduction}\label{sS:vEB}
The essential component of van Emde Boas' data structure is a
reduction that halves the key length $\kl$. Trivially this preserves
that key lengths are powers of two. To do this halving, we would like
to use the reduction above with prefix length $p=\kl/2$. However,
if $\ell$ is too large, we do not have $O(2^p\ell)$ bits of space for
tabulating all prefixes. As a limited start, we can use hashing to tabulate the
above information for all prefixes of keys in $Y$.  Let $U$ be the set of
these prefixes. For all $u\in U$, as above, we store $\fspred_Y[u]$ and
a recursive representation of $\fSuff_Y[u]$. We can then handle all
queries $x$ with a prefix in $U$. However, if the prefix $x_0$ of $x$
is not in $u$, we need a way to compute $\fspred_Y[x_0]$.

To compute $\fspred_Y[x_0]$ for a prefix $x_0$ not in $U$, we
use a recursive representation of $U$. Moreover, with
each $u\in U$, we store the maximal key $\fmax_Y[u]$ in $Y$ with
prefix $u$. Moreover, we define $\fmax_Y[-\infty]=-\infty$. 
We now first compute the predecessor $y_0$ of $x_0$ in $U$, and
then we return $\fmax_Y[y_0]$.

The above reduction spends constant time on halving the key length but
the number of keys may grow in that a key $x=x_0x_1\in Y$ has
$x_0$ in the subproblem $U$ and $x_1$ in the subproblem $\fSuff_Y[x_0]$.
A general solution is that instead of recursing directly on a
subproblem $Z$, we remove the maximal key treating
it separately, thus only recursing on $Z^-=Z\setminus\{\max Z\}$.

In our concrete case, we will consider the reduced recursive
subproblems $\fSuff_Y^-[u]$. We then have the following recursive
pseudo-code for searching the predecessor of $x$:
\begin{quote}
$\fPred(x,Y)$\\
$\null\quad$ $(x_0,x_1)=(\fprefix(x),\fsuffix(x))$\\
$\null\quad$ if $x_0\not\in U$ then return $\fmax_Y[\fPred(x_0,U)]$\\
$\null\quad$ if $x\geq \fmax_Y[x_0]$ then return $\fmax_Y[x_0]$\\
$\null\quad$ $y_1=\fPred(x_1,\fSuff_Y^-[x_0])$\\
$\null\quad$ if $y_1=-\infty$ then return $\fspred_Y[x_0]$\\
$\null\quad$ return $x_0y_1$
\end{quote}
The key lengths have been halved to $\kl'=\kl/2$. We have $n-|U|$ half keys in the suffix 
subproblems $\fSuff_Y^-[x_0]$, and $|U|$ half keys in the prefix
subproblem $U$, so the total number of keys is $n$.

As described above, the space used by the reduction is $O(n\ell)$ bits.

\subsubsection{The final combination}
To solve the predecessor search problem in $O(n2^a\ell)$ bits of space, we will
first tabulate a prefix of length $p\leq\lg n$ as described in
Section \ref{S:pref-tab}, thus reducing the key length to 
$\kl-p\leq 2(\kl-\lg n)$ whish is a power of two. Then we apply the 
van Emde Boas reduction recursively as described in
Section \ref{sS:vEB}, until we get down to a key length below $a$.  This
requires $\lg\frac{\kl-\lg n}a$ recursions. We do not recurse
on empty subproblems. For these we know that the predecessor is
always $0$. Finally we use the complete
tabulation on each subproblem as described in Section \ref{S:tab}.

Since each reduction adds a constant to the search time, the search
time of our solution is $O(\lg\frac{\kl-\lg n}a)$.  The first reduction
uses $O(n\ell)$ bits of space, and the last uses $O(2^a\ell)$ bits of space per
subproblem. Since the subproblems are non-empty, this is $O(n2^a\ell)$ bits of
space in total. Each van Emde Boas recursion uses $O(n\ell)$ bits of space 
where $\ell$ is the current key length. Since $\ell$ is halved each time,
the space of the first iteration with the original key length $\kl$ dominates.
Thus we have proved:
\begin{lemma} Using $O(n2^a\ell)$ bits of space, 
we can represent $n$ $\ell$-bit keys so
to we can search predecessors in $O(\lg\frac{\kl-\lg n}a)$ time.
\end{lemma}

\subsection{Reduction \`{a} la Beame and Fich}\label{sec:real-reduce}
In this section, we will derive better bounds for larger universes
using a reduction very similar to one used by Beame and Fich 
\cite{beame02pred}. Our version of the reduction is captured in
the following proposition:
\begin{proposition}\label{prop:reduce} Let be given an instance of
the static predecessor search problem with $n$ keys of length $\kl$.
Choose integer parameters $q\geq 2$ and $h\geq 2$ where $h$ divides $\kl$.
We can now reduce into subproblems, each of which
is easier in one of two ways:
\begin{description}
\item[length reduced] The key length in the subproblem 
is reduced by a factor $h$ to $\kl/h$, and the subproblem contains
at most half the keys.
\item[cardinality reduced] The number of keys is reduced by a factor 
$q$ to $n/q$.
\end{description}
The reduction costs a constant in the query time. For
some number $m$ determined by the reduction, the reduction uses 
$O((q^{(2h)}+m)\kl)$ bits of space. The total number of keys in the
cardinality reduced subproblems is at most $n-m$, and the total number
of keys in the length reduced subproblems is at most $m$.
\end{proposition}
The original reduction of Beame and Fich \cite[Section
4.2]{beame02pred} is specialized towards their overall quadratic space
solution, and had an assumption that $\kl\leq \kl/h$. They satisfy
this assumption by first applying van Emde Boas' reduction $\lg h$
times. This works fine in their case, but here we consider solutions
to the predecessor search problem where we get down to constant query
time using large space, and then their assumption would be
problematic.

\subsubsection{Larger space}
Recall that we are looking for a solution to the predecessor
problem using $O(n2^a\ell)$ bits of space. In our first simple solution
assumes $a\geq\lg n$ and  $\kl=\omega(\lg n)$. 
With some $h$ to be fixed later we
apply Proposition \ref{prop:reduce} recursively
with $q$ fixed as $2^{a/(2h)}$. Here $a$ and $h$ are assumed
powers of two. Then the bit space used
in each recursive step is $O(2^a\kl)$. Since no subproblem
has more than half the keys, the recursion tree has no
degree 1 nodes. Hence we have at most $n-1$ recursive nodes, so
the total space used in the recursive steps is $O(n2^a\kl)$ bits. 

As described in Section \ref{S:tab}, we can stop recursing when
we get down to key length $a$, so the number of length reductions in
a branch is at most 
$\lg_h \frac wa$. On the other hand, the number of cardinality reductions 
in a branch is at most $\lg_{2^{a/(2h)}} n=\frac{2h(\lg n)}{a}$. 
Thus, for $n\leq s$, the recursion depth is at most
\[\frac{\lg \frac wa}{\lg h}+\frac{2h(\lg n)}{a}.\]
This expression is minimized with 
\[h=\Theta\left(\frac{a\lg \frac wa}{\lg n}/
\lg \frac{a\lg\frac wa}{\lg n} \right)\textnormal,\]
and then we get a query time of  
\[O\left(\frac{\lg \frac wa}{\lg \frac{a\lg \frac wa}{(\lg n)}}\right).\]
Except for the division of $w$ by $a$ in $\frac wa$, the
above bound is equivalent to one anticipated without any proof
or construction in \cite{thorup03stab}. we shall prove that
this bound is tight.

\subsubsection{Smaller space}
We now consider the case where we start with a problem with $\lg n\geq
a/2$ and $\kl=\omega(\lg n)$. We
are now going to appy Proposition \ref{prop:reduce} recursively with a
fixed value of $h$ which is a power of two, but with a changing value
of $q$, stopping when we get a subproblem with only one key, or where
the key length is at most $a$.  While $\lg n\geq a/2$, we use
Proposition \ref{prop:reduce} recursively with $q=\lfloor
n^{1/(4h)}\rfloor$.  However, when we get down to $n\leq 2^{a/2}$
keys, we use $q=2^{a/(4h)}$.
\begin{lemma}\label{lem:small-space} The above construction
uses $O(n2^a\ell)$ bits of space and the search time is 
\[O\left(\frac{\lg \frac {\kl}a}{\lg\frac{\lg \frac {\kl}a}
{\lg\frac{\lg n}{a}}}\right).\]
\end{lemma}
\begin{proof}
First we analyze the search time which is the recursion depth. 
Since we start with key length at most $\kl$ and finish if we
get to $a$, the number of length reductions in a recursion branch is 
$O(\lg_h \frac {\kl}a)$.

For the cardinality reductions, while $n\geq 2^{a/2}$, we note that
it takes less than $4h$ reductions to get from $n$ to $\sqrt n$ keys.
More precisely, in each of these reductions, we have $q>{\sqrt n}^{1/(4h)}$, 
and 
then it takes less than
\[\lg_{\sqrt{n}^{1/(4h)}}\frac{n}{\sqrt n}=\frac{\lg \sqrt n}
{\lg n^{1/(8h)}}=4h\]
cardinality reductions to get down to $\sqrt n$ keys. Thus it
takes $4h$ cardinality reductions to half $\log n$, so to
get from the original value and down to $a/2$, we
need at most $4h\lceil\lg\frac{\lg n}{a/2}\rceil=O(h\lg\frac{\lg n}{a})$
cardinality reductions. In the above argument, we have ignored that
$q$ is rounded down to the nearest integer. However, since
$h$ is a power of two, we can
use the same argument to show that we can have at most $4h$ iterations
while $\log n\in[2^i,2^{i+1})$.

Finally, starting from $n\leq 2^{a/2}$ keys and using
$q=2^{a/(2h)}$, we use at most $h$ cardinality reductions
to get down to a single key. Thus, the total number of
cardinality reducing reductions is $O(h\lg\frac{\lg n}{a})$.
It follows that the total recursion depth is at most
\[O\left(\frac{\lg \frac {\kl}a}{\lg h}+h\lg\frac{\lg n}{a}\right)\]
The search time stated in the lemma is obtained setting
\[h=\frac{\lg \frac {\kl}a}
{\lg\frac{\lg n}{a}}/\lg \frac{\lg \frac {\kl}a}
{\lg\frac{\lg n}{a}}.\]
The bit space bound in each reductive step is $O((q^{(2h)}+m)\kl)$.
We will add up each term separately over the whole recursion tree.

For the $O(m\kl)$ bound, we note that at least $m$ keys get
reduced to length $\kl/h\leq\kl/2$. Thus, the total bit length
of the keys is reduced by at least $m\kl/2$, so we  use $O(1)$
bits per key bit saved. Starting with $n\kl$ key bits, the total space
used is $O(n\kl)$.

Finally, concerning the $O(q^{(2h)}\kl)$ bound,
we have to cases. When $q=\lfloor 2^{(\lg \sqrt n)/(4h)}\rfloor$, the bit
space used is $O(\sqrt n\ell)$. This is $O(\ell/\sqrt n)$ bits of space per
key in the recursion. Following a key $x$ down the branch, we
know that the number of keys is halved in each step, and
this means that the space assigned to $x$ is increased by a
factor $\sqrt 2$. Thus, the total space assigned to $x$ is
dominated by the last recursion, hence $O(\ell)$ bits. Thus, over all
the keys, we get $O(n\ell)$ bits of space for this case.

Finally, when $q=2^{a/(2h)}$, the bit space is $O(2^a\ell)$, and then
the at most $n-1$ recursive nodes give a bit space bound of $O(n2^a\ell)$.
Thus the whole thing adds up to $O(n 2^a\ell)$ bits of space, as desired.
\end{proof}

\subsubsection{Proof of Proposition \latexref{prop:reduce}}
   \label{sec:prop-proof}

In this section, we prove Proposition \ref{prop:reduce}:
\begin{quote}\em
Let be given an instance of
the static predecessor search problem with $n$ keys of length $\kl$.
Choose integer parameters $q\geq 2$ dividing $n$ and $h\geq 2$ dividing $\kl$.
We can now reduce into subproblems, each of which
is easier in one of two ways:
\begin{description}
\item[length reduced] The key length in the subproblem 
is reduced by a factor $h$ to $\kl/h$, and the subproblem contains
at most half the keys.
\item[cardinality reduced] The number of keys is reduced by a factor 
$q$ to $n/q$.
\end{description}
The reduction costs a constant in the query time. For
some number $m$ determined by the reduction, the reduction uses 
$O((q^{(2h)}+m)\kl)$ bits of space. The total number of keys in the
cardinality reduced subproblems is at most $n-m$, and the total number
of keys in the length reduced subproblems is at most $m$.
\end{quote}
In the proof below we will ignore the requirement that a length reduced
subproblem should contain at most half the keys. If one of
these subproblems ends up with two many keys, we can just split
it around the median, adding only a constant to the search time.
 
We will view each key $x$ as a vector $x_1\cdots x_h$ of $h$
characters, each of $c=\kl/h$ bits. We now
provide an alternative to the parallel hashing in
\cite[Lemma 4.1]{beame02pred}. The most significant difference is that
our lemma does not require a word length that is $h$ times bigger than 
$\kl$. Besides, the statement is more directly tuned for our
construction.
\begin{lemma}\label{lem:prefix}
Using $O(q^{2h}\kl)$ bits of space, we can store a set
$Z=\{z^1,...,z^q\}$ of $q$ $h$-character keys so that given a query
key $x$, we can in constant time find the number of whole characters
in the longest common prefix between $x$ and any key in $Z$.
\end{lemma}
\begin{proof}
Andersson et al.~\cite[Section 3]{andersson98sorting} have shown we in
constant time can apply certain universal hash functions $H_1,...,H_h$
in parallel to the characters in a word, provided that the hash values
are no bigger than the characters hashed. Thus, for each $i$
independently, and for any two different characters $x\neq y$, if the
hash values are in $[m]$, then $\Pr[H_i(x)=H_i(y)]\leq 1/m$. Given
$x=x_1\cdots x_h$, we return $H_1(x_1)\cdots H_h (x_h)$ in constant
time. However, the hashed key has the same length as the original
key. More precisely, if the characters have $c$ bits and the hashed
characters are in $[2^b]$, then we have $c-b$ leading zeros in the
representation of $H_i(x_i)$.

We will map each character to $b=2\lg q$ bits. We may here
assume that $b<c$, for otherwise, we can tabulate
all possible keys in $q^{2h}\kl$ bits of space.  For each character
position $i$, we have $q$ characters $z^j_i$, and for random $H_i$ these
are all expected to hash to different values. In particular, we
can choose an $H_i$ without collisions on $\{z^j_i\}_{1\leq j\leq q}$.
Now if $x_i=z^j_i$ we have $H_i(x_i)=H_i(z^j_i)$ and there
is no $z^{j'}_i\neq z^j_i$ with $H_i(x_i)=H_i(z^{j'}_i)$.

Next, consider the set $A$ of values $H_1(x_1)\cdots H_h(x_h)$ over
all possible vectors $x=x_1\cdots x_h$. These vectors are $ch$ long,
but since only the $b$ least significant bits are used for the hash
values of each character, there are at most $2^{bh}$ different values
in $A$. Using the linear space 2-level hashing of Fredman et
al.~\cite{fredman84dict}, we construct a hash table $\mathcal H$ over
$A$ using $O(2^{bh}\kl)$ bits of space.  With the entry ${\mathcal
H}(H_1(x_1)\cdots H_h(x_h))$, we store the key $z^j$ so that
$H_1(z^j_1)\cdots H_h(z^j_h)$ has the longest possible prefix with
$H_1(x_1)\cdots H_h(x_h)$. The key $z^j$ is found from $x$ in constant
time.

We now claim that no key $z^{j'}$ can agree with $x$ in more characters
than $z^{j'}$. Suppose for a contradiction that $z^{j}$ agrees with
$x$ in the first $r-1$ characters but not in character $r$, and that
$z^{j'}$ agrees in the first $r$ characters. Then
$H_1(x_1)\cdots H_r(x_{r})=H_1(z^{j'}_1)\cdots H_r(z^{j'}_{r})$. However,
since $H_r$ is 1-1 on $\{z^j_r\}_{1\leq j\leq q}$, $H_r(z^{j}_{r})\neq
H_r(z^{j'}_{r})=H_r(x_{r})$.

All that remains is to compute the number of whole characters in the
common prefix of $x$ and $z^j$. This can be done by clever use of
multiplication as described in \cite{fredman93fusion}. A more
practical solution based on converting integers and to floating point
numbers and extracting the exponent is discussed in
\cite{thorup00prique}.
\end{proof}

Using Lemma \ref{lem:prefix}, we can compute in constant time the 
longest common prefix, $\fcompref_Z[x]$, in whole characters, between $x$ and 
any key in $Z$.
Also, if $x\not\in Z$, we can get the prefix $\fcompref^{\,+}_Z[x]$
that has one more character from $x$.

We are now return to the proof of Proposition \ref{prop:reduce} which
is similar to the one in \cite[Section 4]{beame02pred}. Out of our original
set $Y$ of $n$ keys, we pick a subset $Z=\{z^1,...,z^q\}$ of $q$ keys
so that there is a key from $Z$ among any sequence of $\lceil
n/q\rceil$ consecutive keys from $Y$. We apply Lemma \ref{lem:prefix}
to $Z$. Thereby we use $O(q^{2h}\kl)$ bits of space. 
We are going to consider two types of subproblems.

\paragraph{Cardinality reduced problems}
First we have the cardinality reduced subproblems. These are of the following
type: we take a key from $Y\setminus Z$ and consider the prefix
$v=\fcompref^{\,+}(y,Z)$. Let $\fAgree_Y[v]$ denote the keys from $Y$
that have prefix $v$. These keys are consecutive and they do not 
contain any key from $Z$, so $|\fAgree_Y[v]|<q$. 
We use 2-level hash table for the prefixes in 
$V=\{\fcompref^{\,+}(y,Z)\,|\,y\in Y\setminus Z\}$. With $v\in V$,
we store $\fpred_Y[v]$ and $\fmax_Y[v]$ as defined in the previous 
section, that is, $\fpred_Y[v]$ is the strict predecessor in $Y$ of $v$ 
suffixed by zeros, and $\fmax_Y[v]$ is the largest key in $Y$ with
prefix $v$. Finally, as the cardinality reduced subproblem, we
have $\fAgree^-_Y[v]
=\fAgree_Y[v]\setminus \{\max \fAgree_Y[v]\}$. 

This above informatoin suffices to find the predecessor of any query
key $x$ with $\fcompref^{\,+}(x,Z)=v$. The bit space used above is $O(|V|\kl)$.
Each cardinality reduced subproblem $\fAgree^-_Y[v]|$ has
at most $q-2$ keys, and they add up to a total of $n-|Z|-|V|$ keys.

\paragraph{Length reduced subproblems}
For query keys $x$ with $\fcompref^{\,+}(x,Z)\not\in V$, we will consult
length reduced subproblems defined over the set $U$ of prefixes of
keys in $Z$. We will have a 2-level hash table over $U$. For each
$u\in U$, let $\fNextchar_Y[u]$ be the set of characters $c$ such that
$uc$ is a prefix of a key in $Y$. We will have a length reduced
subproblem over the characters in $\fNextchar^-_Y[u]=\fNextchar_Y[u]
\setminus\{\max \fNextchar^-_Y[u]\}$. As complimentary information, we
store $\fpred_Y[v]$ and $\fmax_Y[v]$.

Now, consider a query key $x$ with $\fcompref^{\,+}(x,Z)\not\in V$.
Let $u=\fcompref(x,Z)$ and let $d$ be the subsequent character in
$x$, that is, $ud=\fcompref^{\,+}(x,Z)$. Then $d\not\in \fNextchar_Y[u]$. 
Suppose $x$ is between the smallest and the largest key in $Y$ with
prefix $u$. If $c$ is the predecessor of $d$ in $\fNextchar^-_Y[u]$, 
then the predecessor of $x$ 
in $Y$ is the largest key with prefix $uc$. However, $uc\in V$, so
the predecessor of $x$ is the $\fmax_Y[uc]$ stored under the length 
reduced subproblems. 

The above length reduction used $O(\ell)$ bits for each $u\in U$ and
$c\in \fNextchar_Y[u]$. Consider $c\in
\fNextchar_Y[u]$. There can be at most $U$ cases where $uc\in
U$. Otherwise, we have $uc=\fcompref^{\,+}(y,Z)\in V$. The total bit space of
the length reduction is hence $O((|U|+|V|)\kl)$.

We will now prove that the total number of keys in the
length reduced subproblems $\fNextchar^-_Y[u]$ is at most $|Z|+|V|$.
Above we saw that if a character $c\in \fNextchar_Y[u]$ did not represented
a prefix in $U$, it represented a prefix in $V$. Those
representing prefixes in $U$ can also be viewed as representing
children in the trie over $Z$. The total number of such
children is at most $|Z|$ plus the number of internal trie nodes,
and since we for $\fNextchar^-_Y[u]$ subtracted a node for each internal
trie node $u$, we conclude that the total number of keys is
the length reduced subproblems is bounded by $|Z|+|V|$.

\paragraph{Pseudo-code} We now have 
the following recursive pseudo-code for searching the 
predecessor of $x$ in $Y$:
\begin{quote}
$\fPred(x,Y)$\\
$\null\quad$ if $x\in Z$ return $x$\\
$\null\quad$ let $ud=\fcompref^{\,+}_Z(x)$ with $d$ the last character\\
$\null\quad$ if $ud\in V$ then\\
$\null\quad\quad$ if $x\geq \fmax_Y[ud]$ then return $\fmax_Y[ud])$\\
$\null\quad\quad$ $y=\fPred(x,\fAgree_Y^-[uc])$\\
$\null\quad\quad$ if $y=-\infty$ then return $\fspred_Y[ud]$\\
$\null\quad\quad$ return $y$\\
$\null\quad$ if $x\geq \fmax_Y[u]$ then return $\fmax_Y[u]$\\
$\null\quad$ $c=\fPred_Y(d,\fNextchar_Y^-[u])$\\
$\null\quad$ if $c=-\infty$ then return $\fspred_Y[u]$\\
$\null\quad$ return $\fmax_Y[uc]$
\end{quote}

\paragraph{Final analysis}
This almost finishes the proof. Let $m=|Z|+|V|$. Then
we have at most $n-m$ keys in cardinality reduced subproblems
and at most $m$ keys in length reduced subproblems.

The total bit space used is $O(q^{2h}\kl)$ for the implication of 
Lemma \ref{lem:prefix}, $O(|V|\ell)$ for the cardinality reduction, and
$O((|U|+|V|)\kl)$ for the length reduction. Here $O(|U|)=O(hq)=O(q^{2h})$ and
$|V|<m$, so the total bit space is $O((q^{2h}+m)\kl)$.  {\bf This completes
the proof of Proposition \ref{prop:reduce}.}\hfill \qed

\bibliographystyle{plain} 
\bibliography{../../general}

\end{document}